\documentclass{JINST}
\pdfoutput=1 
\usepackage{ifpdf}

\title{Response of Multi-strip Multi-gap Resistive Plate Chamber}

\author{Ushasi Datta$^a$ \thanks{Corresponding author.}~, S. Chakraborty$^a$,  A. Rahaman$^a$, P. Basu$^a$, J.Basu$^a$, 
D. Bemmerer$^b$, K. Boretzky$^c$ , Z. Elekes$^b$,  M. Kempe$^b$,G. Munzenberg$^c$, H. Simon$^c$, M. Sobiella$^b$,
 D. Stach$^b$, A. Wagner$^b$, D. Yakorev$^b$ \\
\llap{$^a$}Nuclear Physics Division, Saha Institute of Nuclear Physics,\\
  1/AF Bidhannagar, Kolkata 700064, India\\
\llap{$^b$}Institute for Radiation Physics, Forschungszentrum Dresden-Rossendorf,\\
  Bautzner Landstra\ss e 400, 01328 Dresden, Germany\\
\llap{$^c$}Gesellschaft f\"{u}r Schwerionenforschung, Darmstadt, 64291, Germany\\

  E-mail: \email{ushasi.dattapramanik@saha.ac.in}}

\abstract{A prototype of Multi-strip  Multi-gap Resistive Plate chamber (MMRPC) with active area
 40 cm $\times$ 20 cm has been developed at SINP, Kolkata. Detailed response of the developed 
detector was studied with the pulsed electron beam from ELBE at Helmholtz-Zentrum Dresden-Rossendorf.
 In this report the response of SINP developed MMRPC with  different controlling parameters is
 described in details. The obtained  time resolution ($\sigma_t$) of the detector after slew 
correction was 91.5$ \pm $3 ps. Position resolution measured along ($\sigma_x$) and
across ($\sigma_y$) the strip was 2.8$\pm$0.6 cm and 0.58 cm, respectively. The measured 
absolute efficiency of the detector for minimum  ionizing particle like electron was
 95.8$\pm$1.3 $\%$. Better timing resolution of the detector can be achieved by restricting 
the events to  a single strip. The response of the detector was mainly in avalanche mode but a 
few percentage of streamer mode response  was also observed. A comparison of the response of these
 two modes with trigger rate was studied.}


\begin{document}

\section{Introduction}

 Three decades  ago the Resistive Plate Chamber (RPC) \cite{san81,car88} was invented to overcome  
several problems of parallel plate chambers. Unlike parallel plate chambers, electrodes of
 RPCs  are made of resistive material like Bakelite or glass. This has the effect that only 
a  limited  part of the electrode  is discharged  during the passage of an  ionizing particle 
with subsequent  avalanches or streamers, while the rest of the electrode remains unchanged. 
To improve timing resolution, Multi-gap Resistive Plate Chamber \cite{car96,pfonte}  (MRPC) is an 
intelligent modification of an RPC by increasing the electric field across the gap
 and creating thinner layers of gas gap by inserting (electro-statically) floating glasses
 between anode and cathode. Time resolution better than 50 ps $\sigma$ at 99$\%$ for Minimum Ionizing
 Particle (MIPs) can be achieved by operating a MRPC detector in  avalanche mode. Further, 
segmented structure in readout strips design makes the position resolution 
 of MRPC as good as 0.5 mm\cite{yjin}. MRPCs are, thus, high granularity, high-resolution 
inexpensive TOF system (compared to standard scintillator with PMTs) appropriate for large scale 
applications. This includes both fundamental research in particle physics large scale experiments  
i.e, STAR\cite{Ammos,bonn}, HARP\cite{Barra},
 ALICE\cite{Akin,Akin1}, astrophysics \cite{cam12},  cosmology \cite{upp},  nuclear physics HADES\cite{Belver,Belver1},
 FOPI\cite{Schuettauf, Schuettauf1,Petrovici}, and applied research in medical imaging 
(cost effective Positron Emission Tomography) \cite{Blanco,dor}, security purpose like cosmic muon tomography
 \cite{Boroz}, climate change\cite{bal}  etc.  The   response   of  such detector under irradiation by the $\gamma$-ray 
 \cite{datt} and neutrons has not been studied in details \cite{jam,yan,datt,zol}. In order to explore 
in this direction a prototype of Multi-strip Multi-gap Resistive Plate Chamber (MMRPC) has been developed at SINP,
 Kolkata. In first stage of development, the design  was focused  on the feasibility study  of  
the MRPC as an active part of high energy, high efficiency neutron TOF. 
The response of the   MMRPC detector using $\gamma$ and cosmic muon  was extensively studied at SINP, 
laboratory , \cite{datt}. Later,  the detector was taken to the electron linac ELBE at 
Helmholtz-Zentrum Dresden-Rossendorf, Germany to  study its electron response. The optimum operating condition 
(w.r.t efficiency, time  resolution, position resolution, etc.)
 was studied.   In the following, the response of our newly developed MMRPC detector for electrons will be 
discussed in details.  

\section{ Detector construction and design details}\label{sec2}
As a first step, we have developed a  prototype of double stack, four gas gap  glass MMRPC of 
size  40 cm $\times$20 $cm$ with segmented anode  strip .  Figure \ref{picture of mmrpc cross_view}
shows the cross-section view of the developed MMRPC. The anode strip is made of PCB
 with thin layer of gold coating and has nine anode read out strips with two strips at the
 edges grounded via 50 ohm resistance.   This  was designed by a group of scientists  
at SINP and manufactured by local workshop of Kolkata.  The    anode
 strip size is 2.0 cm wide and 40 cm long. Each 
strip are read out at both the end. Thus for this MMRPC, in total eighteen  
readout-signals   are there.   Figure \ref{picture of anode strip} shows 
the anode strips of the MMRPC detector. 
The cathode plates are made by introducing a thin layer of conducting material 
on 1 mm thick float glass.  Negative high voltage is  supplied  to the cathode 
whereas the anode is kept at the ground potential.  
Four uniform and symmetric gas gaps  are made between two cathode plates. 
Fishing lines ($0.3$ mm diameter) are used as spacer between glass plates. 
Thus, two sets of 1 mm thick floating glass  and 0.3 mm gas gap act as 
dielectric material between the cathode and anode plates.  This detector 
structure is housed in an  aluminum chamber which was  designed  and build at 
workshop of  SINP, kolkata \cite{datt}.
 A custom built gas system{\cite{Jos09}  was used where four gases can be 
mixed and delivered to four RPCs. A moisture  meter is also  available for
 monitoring  of the gases. The gas system with the mass flow controller produce 
 a gaseous mixture of   R134a (C$_2$H$_2$F$_4$),   Sulfur hexafluoride (SF$_6$), 
Isobutene (ISO-C$_4$H$_{10}$) in the ratio of 85:6.3:8.5 at normal atmospheric
 pressure condition used as the counting gas within the MMRPC detector. 
The detector is flushed with 
the gas mixture in  every $\sim$ 8 hrs.

\subsection{Experimental Set-up}
 Electron response of the developed MMRPC detector was studied at the radiation physics cave of 
ELBE, Dresden, Germany
\begin{figure}[ht!]
\begin{center}
\includegraphics[width=0.7\textwidth]{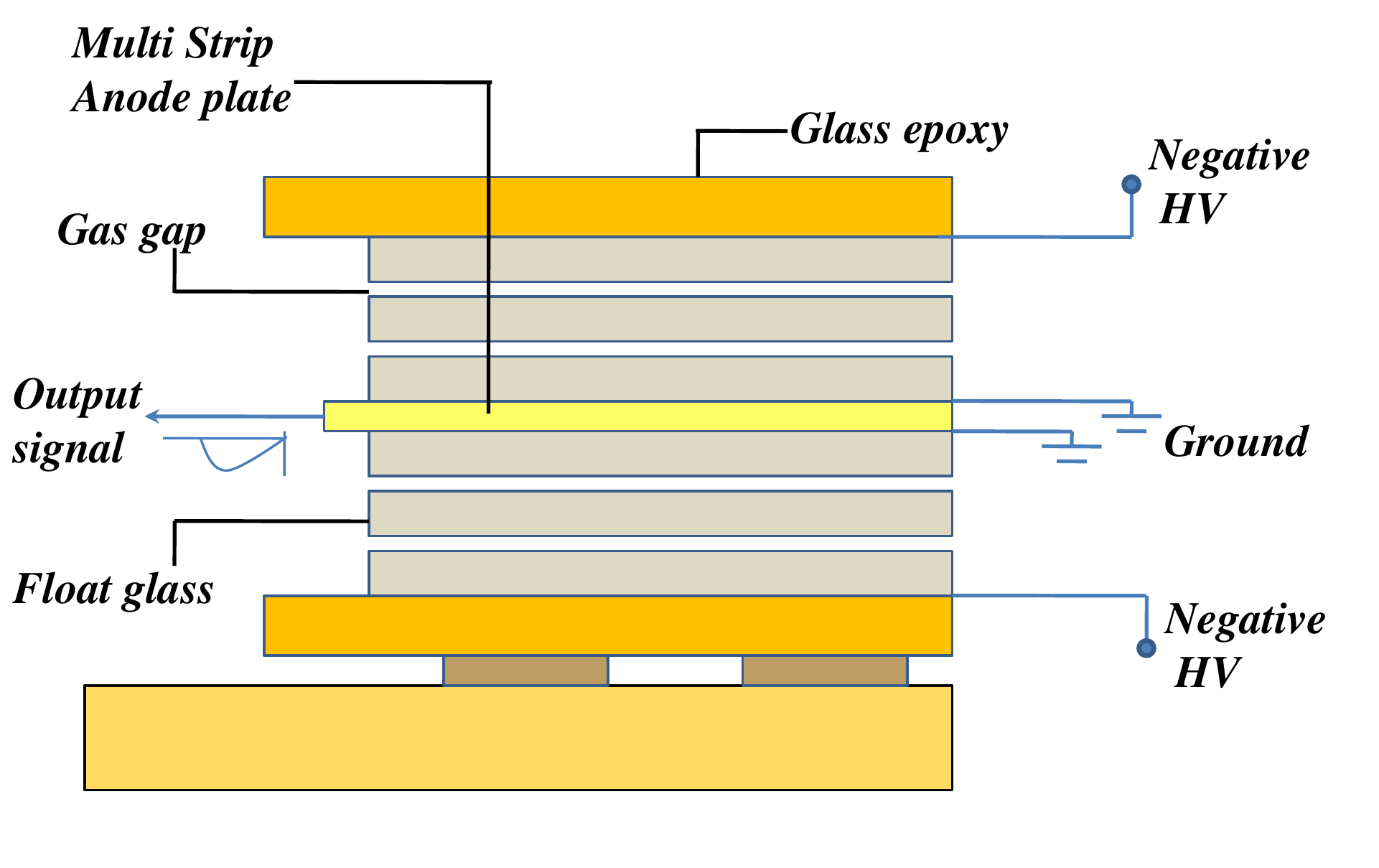}
\end{center}
\caption{ Cross-section view of developed Multi-strip Multi-gap Resistive Plate Chamber.}
\label{picture of mmrpc cross_view}
\end{figure}

\begin{figure}[ht!]
\begin{center}
\includegraphics[width=0.7\textwidth]{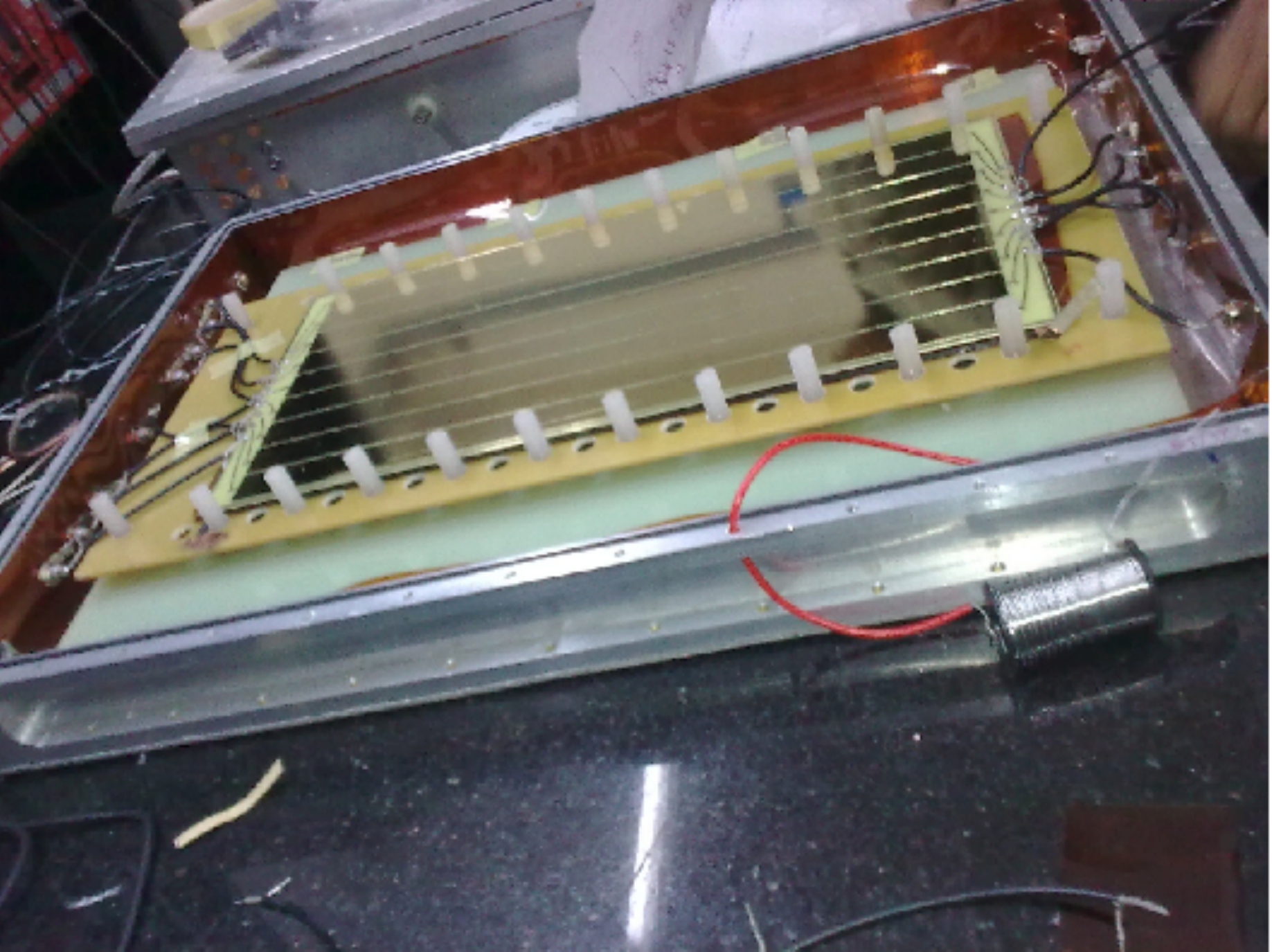}
\end{center}
\caption{ Photograph of Multi-strip Multi-gap Resistive Plate Chamber with segmented anode strip.}
\label{picture of anode strip}
\end{figure}

\begin{figure}[ht!]
\begin{center}
\includegraphics[width=0.7\columnwidth]{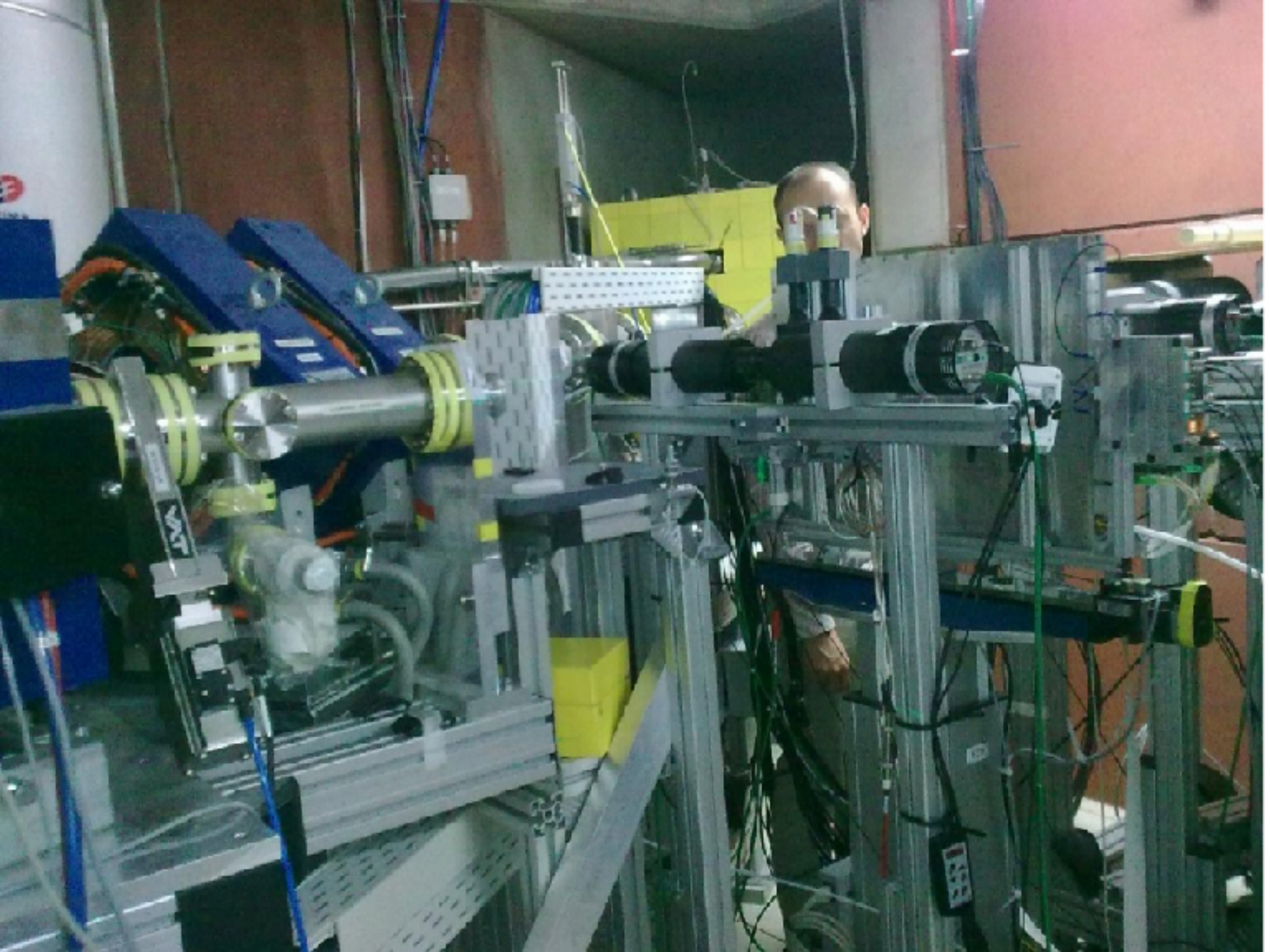}
\end{center}
\caption{ Photograph of the experimental set-up for studying the electron response of MMRPC at ELBE.}
\label{photograph of experimental set-up}
\end{figure} 

For detailed  study of  the electron   response of the SINP  prototype  the detector was irradiated  with  the   electron beam of 
 29  MeV from Electron Linac with high Brilliance and low Emittance (ELBE) facility at Helmholtz-Zentrum Dresden-Rossendorf,  
Dresden,  Germany. The radiation source ELBE is based on a super conducting electron 
linac delivering electron beams in the energy range of 10 - 40 MeV. The accelerator produces a quasi-continuous electron beam with 
micro-pulse repetitive rate ranging from 1Hz to 260 MHz.  At a beam current of 1 mA the charge of electron bunch can be varied 
from 1 fC up to 77 pC by changing the source current at the pulse injector i.e. the 250 kV electron gun. Four cavities are operated 
in a nominal accelerating gradient of 10 MeV/m (10 MeV energy gain per cavity). One or more special aluminum foils are introduced in 
the optical path between two accelerator cavities.  This leads to the production of single electron bunch of pulse width less than
 10 ps\cite{neu}. This  short micro-pulse duration makes it very attractive to use the RF signal of ELBE as time reference for time measurement.

\begin{figure}[ht!]
\begin{center}
\includegraphics[width=0.7\columnwidth]{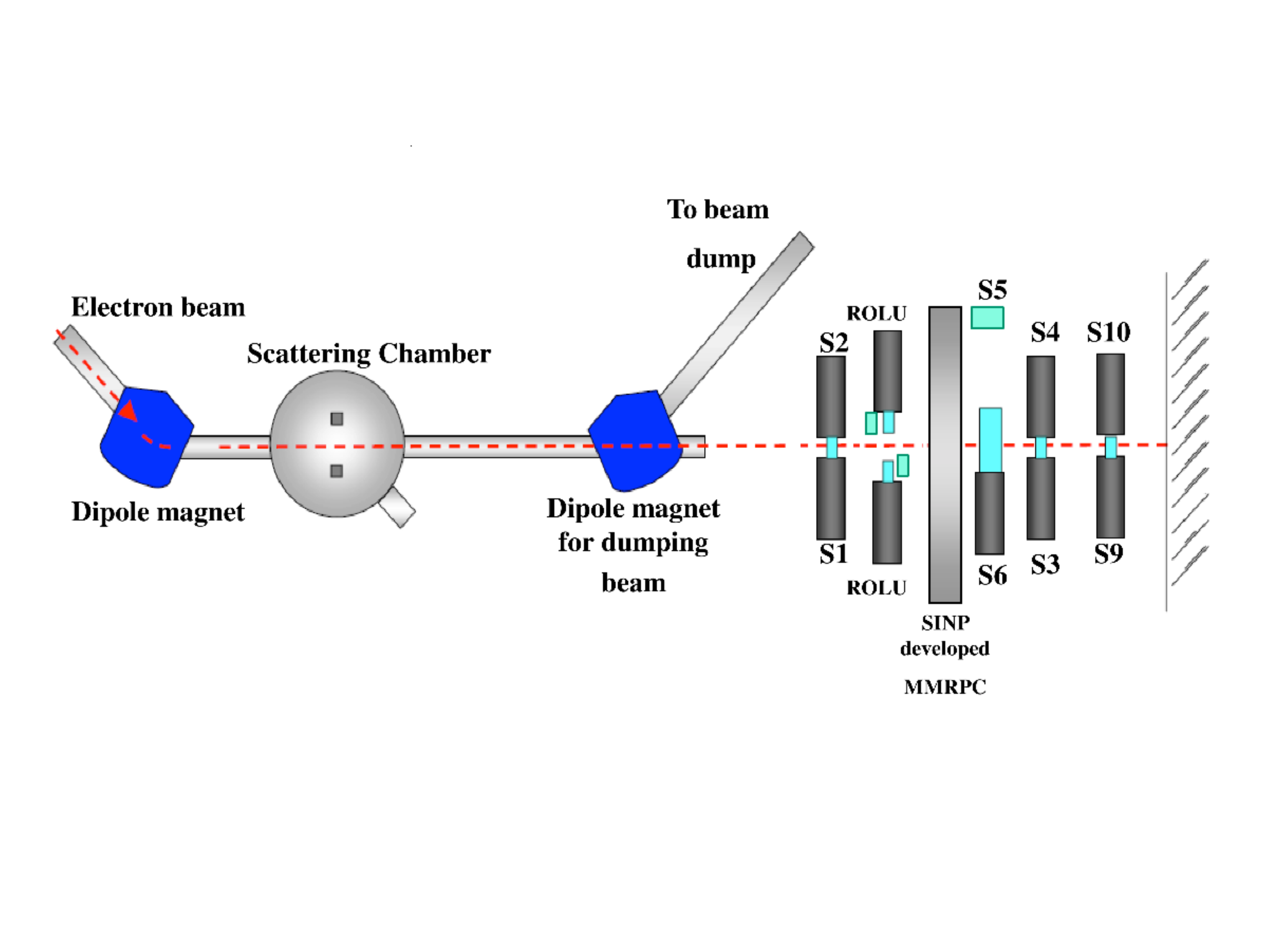}
\end{center}
\caption{Block diagram of the experimental setup for studying the electron response of MMRPC. 
The pulsed electron beam, shown by dashed line,  passes through the  air column of  41 cm and then bombards on the set-up. }
\label{MMRPC using ELBE}

\end{figure}
\begin{figure}[h!]
\begin{center}
\includegraphics[width=0.7\columnwidth]{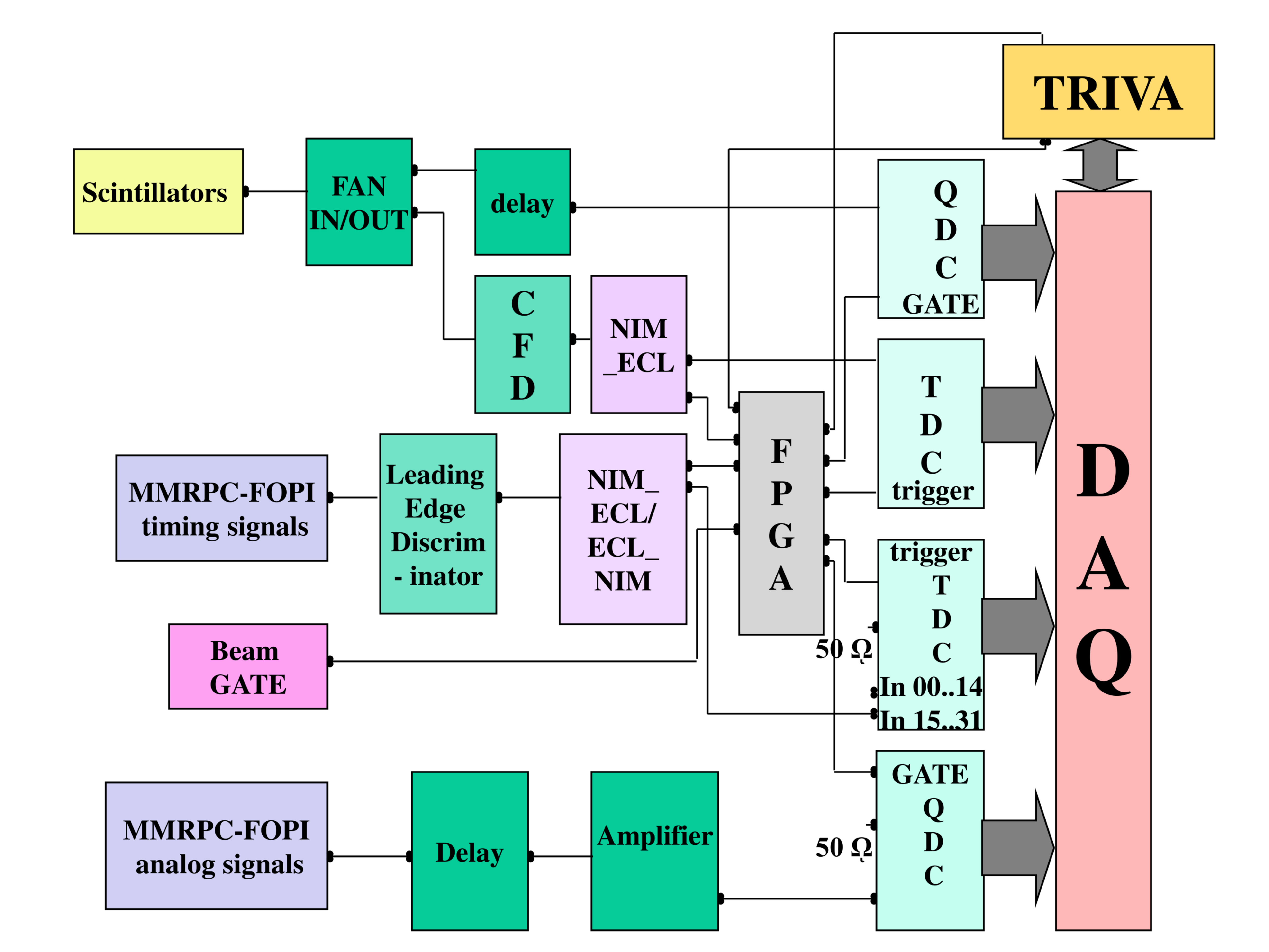}
\end{center}
\caption{Block diagram of the electronics used for the processing the timing and analog signals of MMRPC and scintillators.}
\label{signals of MMRPC}
\end{figure}

The electron energy was  chosen to be 29 MeV with pulse width less than 10 ps. The beam was operated in single-electron per bunch
 mode \cite{neu} and was allowed to impinge on   the test set-up through a 40 micrometer thick Beryllium window. Figure 
\ref{photograph of experimental set-up} shows a photograph of the experimental set-up at the radiation physics Cave at ELBE during 
this experiment. The geometry of the experimental set-up is shown in Figure \ref{MMRPC using ELBE}. The pulsed electron beam 
(shown by dashed line in figure 3) passes through the  air column of  41 cm  before  bombarding the set-up.  The setup consists of  
several scintillators on the path of electron beam.  Three scintillators (BICRON BC-408)   were read out by a pair of 2 inch Photo
 Multiplier Tubes (XP2020) hence delivering signals from S1/S2, S3/S4 and S9/S10.  In between the scintillator S1/S2 and  SINP MMRPC 
detector (Figure 2 and 3) there was an active collimator ( ROLU ) which consisted of four plastic scintillators, each read out by a PMT.
 There was another plastic scintillator S6,  placed between MMRPC and S3/S4. The beam profile was obtained  using a plastic scintillator
  S5  placed after  the MMRPC detector. Both MMRPC and this plastic scintillator can be moved using a remote controlled motor to scan both
 in horizontal and in vertical direction.  The time information of the scintillators were extracted using constant fraction discriminator 
(ORTEC CF 8000).  FOPI cards\cite{cio} were used as the Front End Electronics (FEE) of  the  MMRPC. The timing signals from the FEE
 card were shaped  using leading edge discriminator (Philips octal 
Discriminator 710). A logical AND of the timing signals from the scintillators S1, S2, S6 and the RF provides  the master trigger logic 
for the  DAQ system.  The MMRPC analog signals after being delayed (125 ns) through Aircell 5 patch delay were digitized with a CAEN QDC 
V965. The digitized  amplitude signals were used for the  time-slewing (walk) correction. The essential component of the DAQ is a
 CAEN 32 channel multi-hit TDC of type (V1290A) with a least significant bit (LSB) size of 25 ps.  The Multi-Branch System 
(MBS)\cite{mbs} was used for the data acquisition (DAQ).   Figure \ref{signals of MMRPC}  shows the block diagram of the electronics
 used for the processing and acquiring the signals from the MMRPC and the scintillators detectors.
 
 \begin{figure}[ht!]
\begin{center}
\includegraphics[width=0.7\textwidth]{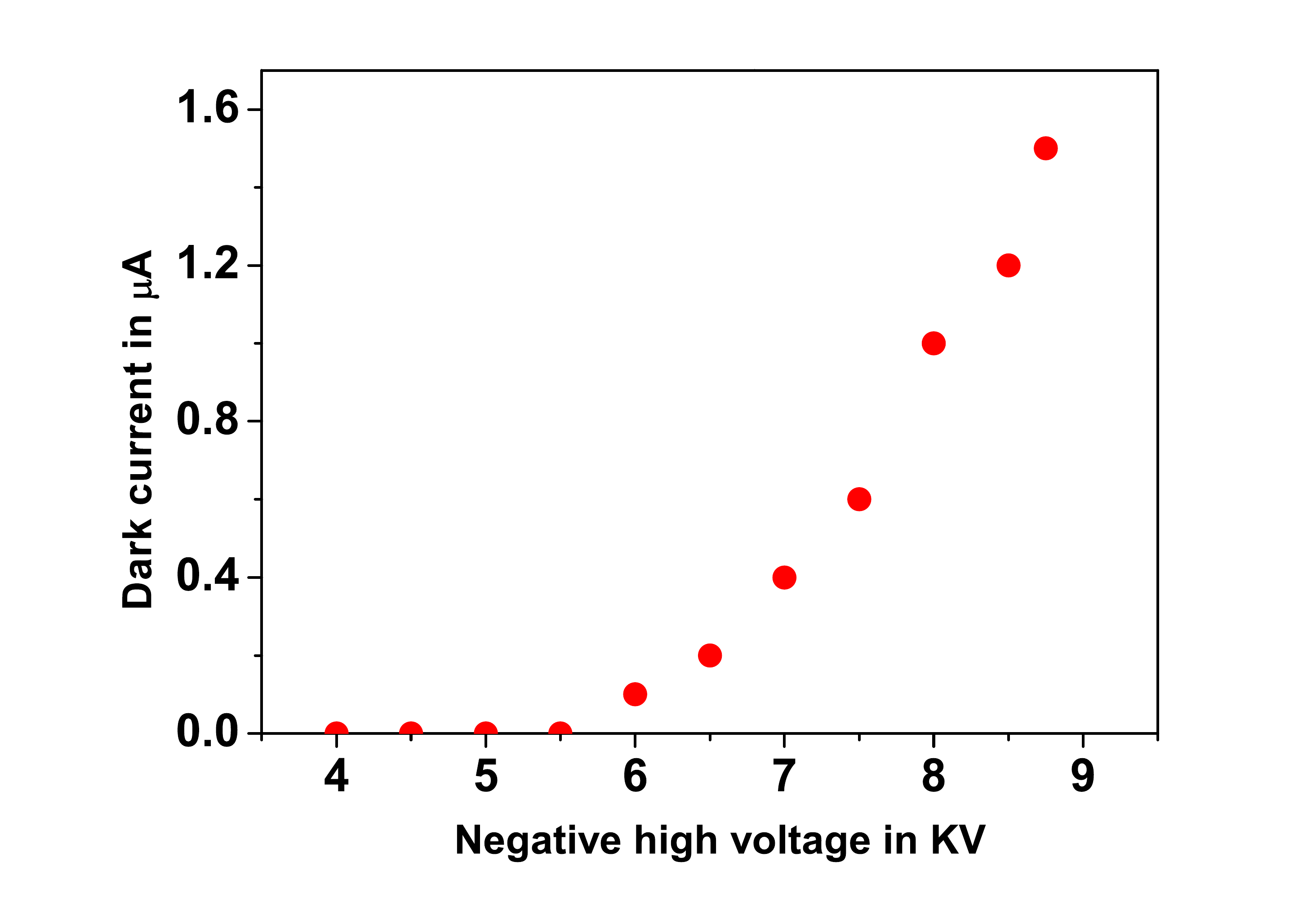}
\end{center}
\caption{Plot of dark current against bias voltage of MMRPC.}
\label{dark current}
\end{figure}

\begin{figure}[h!]
\begin{center}
\includegraphics[width=0.7\textwidth]{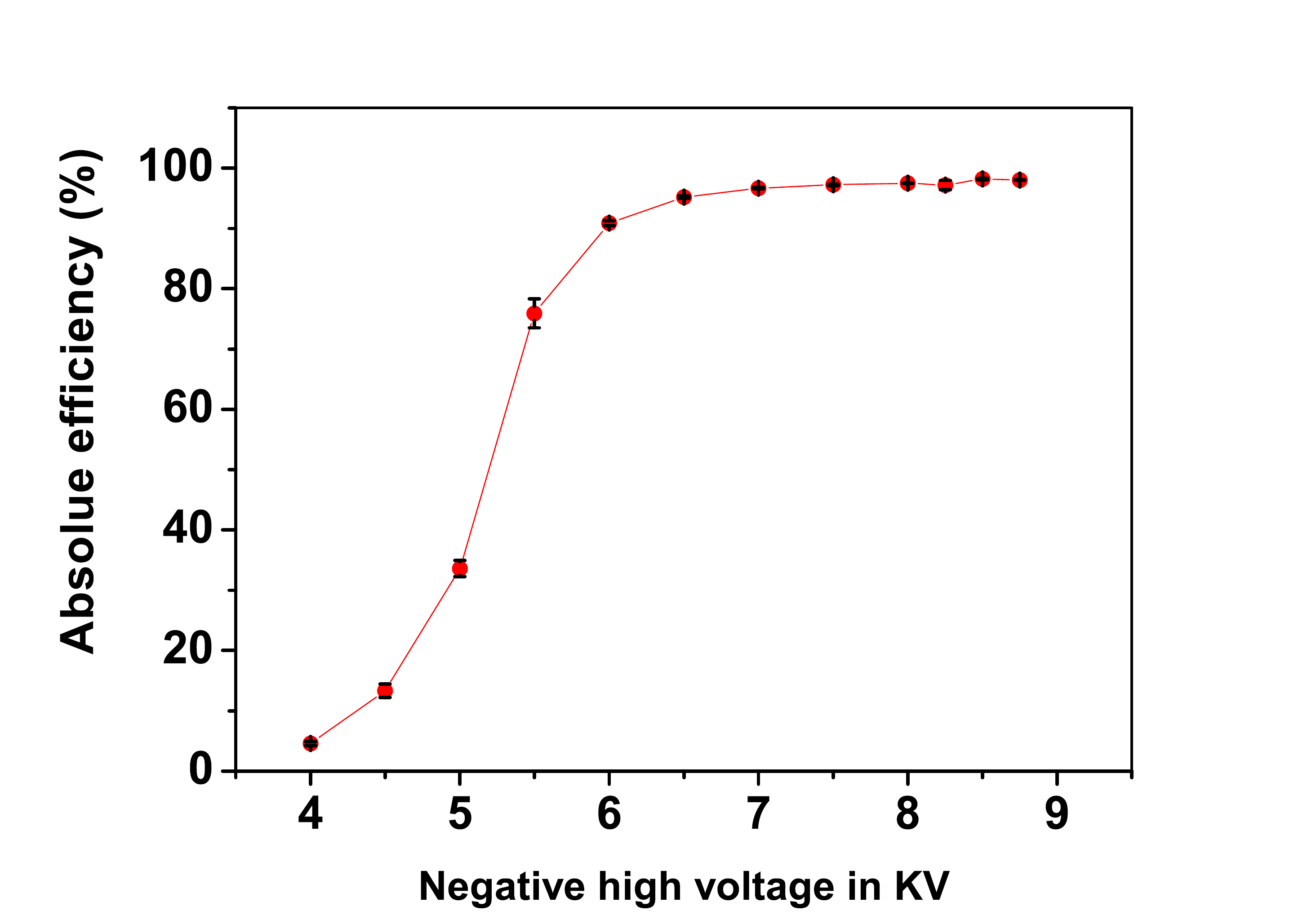}
\end{center}
\caption{Plot of absolute efficiency against bias voltage of MMRPC.}
\label{efficiency}
\end{figure}   

To find the optimum operational condition of the prototype, the detector efficiency and time resolution were studied  as a 
function  of the applied high voltage using a beam with a flux of 50 Hz/cm$^2$. During this procedure the beam was  focused on a single
  strip of MMRPC. In subsequent steps, the same procedure was repeated  with increased trigger rate of the beam   up to  1000 Hz in steps   of 200Hz. 
Measurements were  also performed with the beam spot focused on each of the other individual strips of MMRPC. All these measurements  
 were performed with the beam spot positioned at the middle along the length of individual strips. The beam spot  was further moved along the
 strip to the edges for making position measurements.    

\begin{figure}[ht!]
\begin{center}
\includegraphics[width=0.7\textwidth]{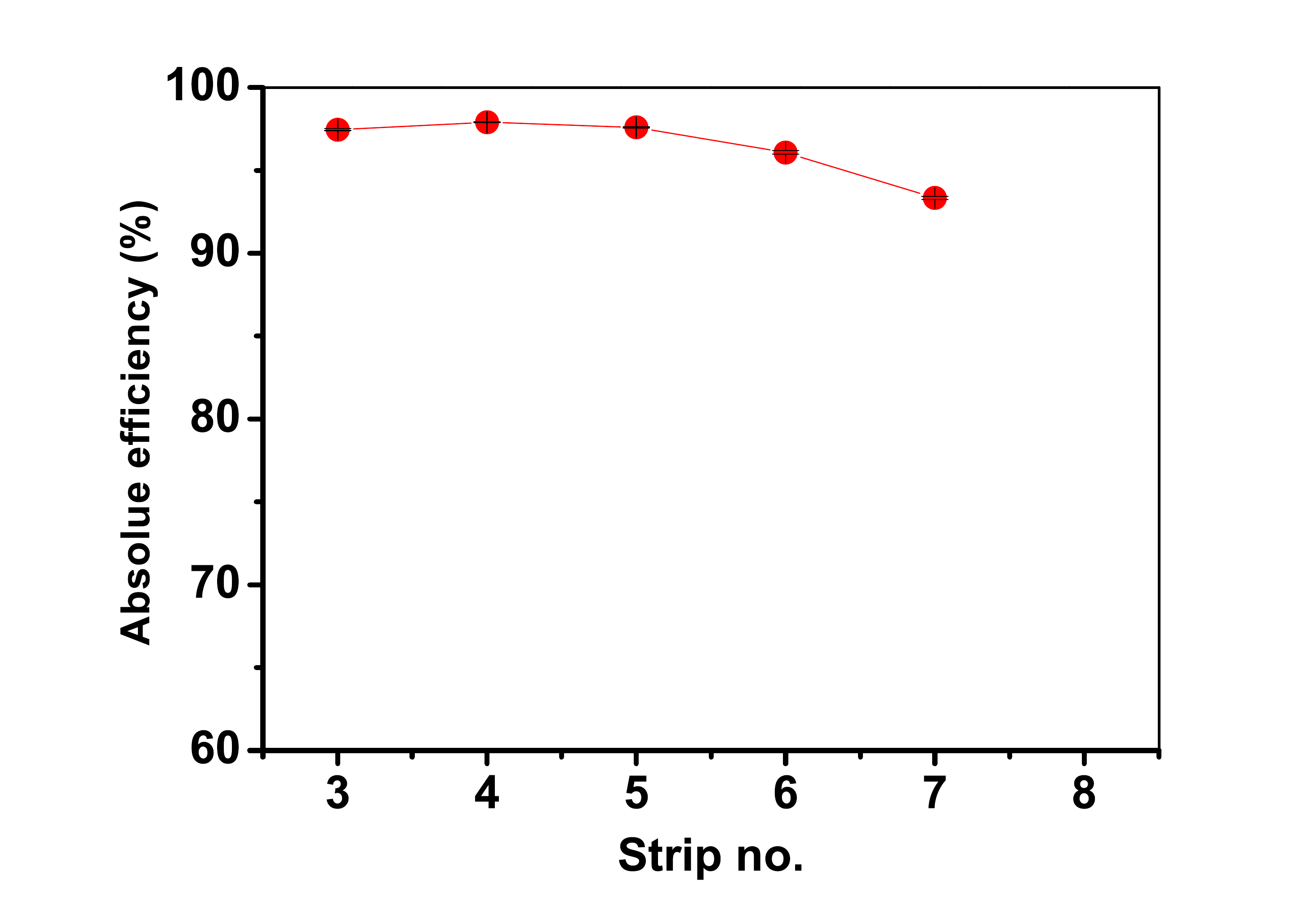}
\end{center}
\caption{Variation of absolute efficiency against the strip number of MMRPC.}
\label{efficiency against strip}
\end{figure}

\begin{figure}[h!]
\begin{center}
\includegraphics[width=0.7\textwidth]{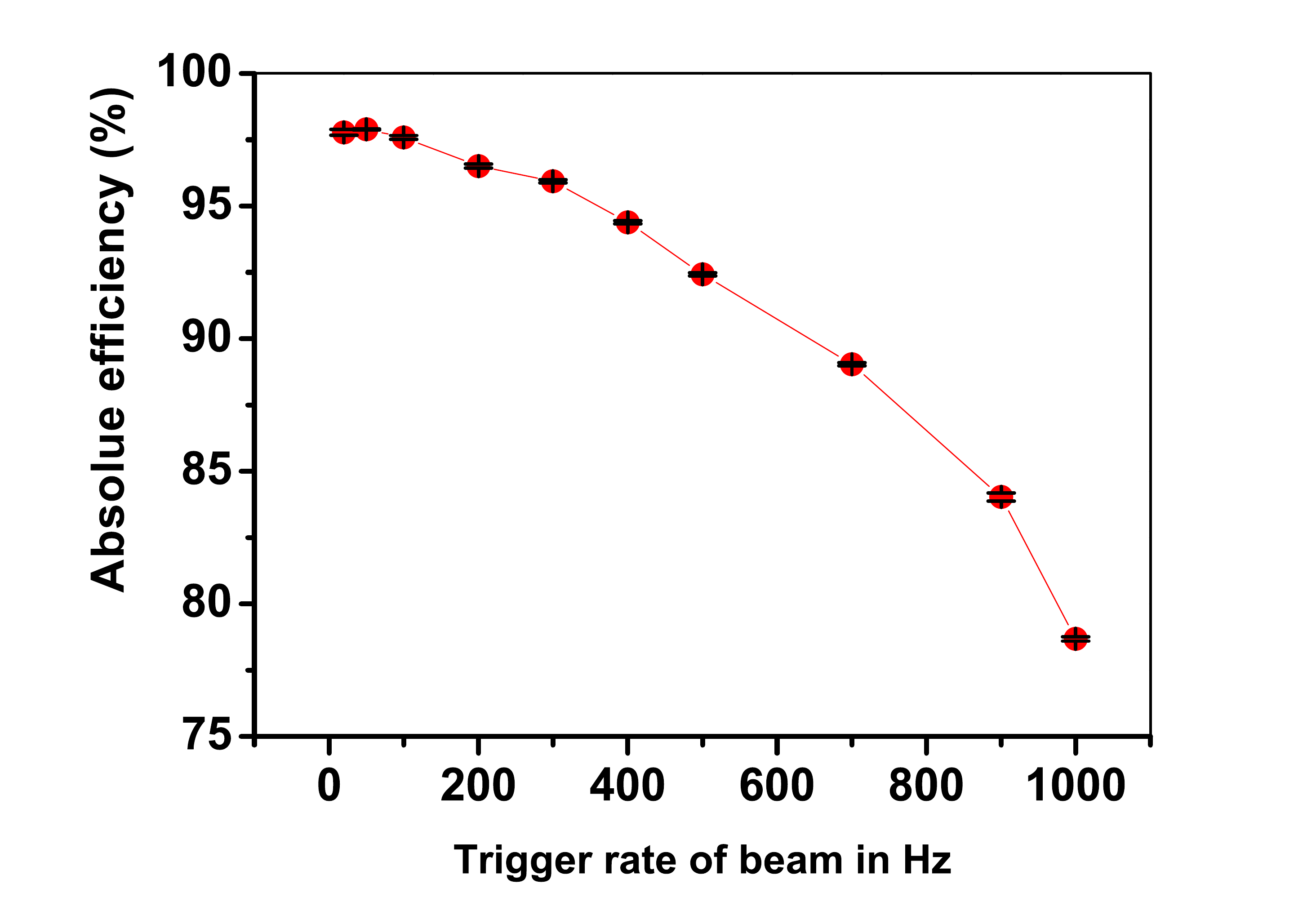}
\end{center}
\caption{Variation of absolute efficiency against the trigger rate for fixed bias voltage.}
\label{efficiency against trigger}
\end{figure}

%%%%%%%%%%%%%%%%%%%%%%%%%%%%%%%%%%%%%%%%%%%%%%%%%%%%%%%%%%%%%%%%%%%%%%%%%%%%%%%%%%%%%%%%%%%%%%%%%%%%%%%%%%%%%
\section{ Analysis and Results}
%\label{sec4}

Figure \ref{dark current} shows a plot of dark current against bias voltage.
The beam profile was scanned with the help of scintillator detector mounted on a linear step motor drive.  The scintillator 
detector was moved horizontally to identify the center of  the beam.

A vertical scan was performed keeping the horizontal position of the detector fixed at beam center. The measured FWHM of the beam was $\sim$2.5 cm.
The  beam profile being larger than the strip width, absolute efficiency of the MMRPC detector was determined using the following logic. 
 The beam spot was focused on strip number 4. A valid event or ``good events'' of the particular strip of MMRPC was considered 
when a strip had a valid event from the TDCs at  the two ends of that particular  anode strip. Now, the dimension of the beam spot
 clearly shows that any one of neighboring three strips can be  fired for a 
particular event. Hence, ``good events'' are those which have good signals either in   strip 4  or its neighboring strip (i.e, strip 3 and 5). 
Hence, the absolute efficiency of our developed MMRPC can be defined as the ratio of the number of ``good events ''  or valid events of MMRPC to 
number of events triggered in that particular time interval. 
 Figure \ref{efficiency}, \ref{efficiency against strip} and \ref{efficiency against trigger} show the plot of the absolute efficiency against the bias 
voltage of MMRPC, strip number of MMRPC and triggering rate of the incident beam, respectively. In this analysis, the absolute  efficiency of MMRPC 
 reaches 95.8$\pm$1.3 $\%$ at bias voltage greater than 6.5 KV. A sufficiently long plateau has been observed allowing for  safe detector operation. 
Therefore, a high voltage of 7.5 kV was applied  to the Cathode of the detector  for further investigations. It is also visible that efficiency of 
the detector decreases  with increasing trigger rate. In the plot of the absolute efficiency against strip number  (Figure \ref{efficiency against strip})
 almost an uniform efficiency of the detector was observed.  
\begin{figure}[h!]
\begin{center}
\includegraphics[width=0.7\textwidth]{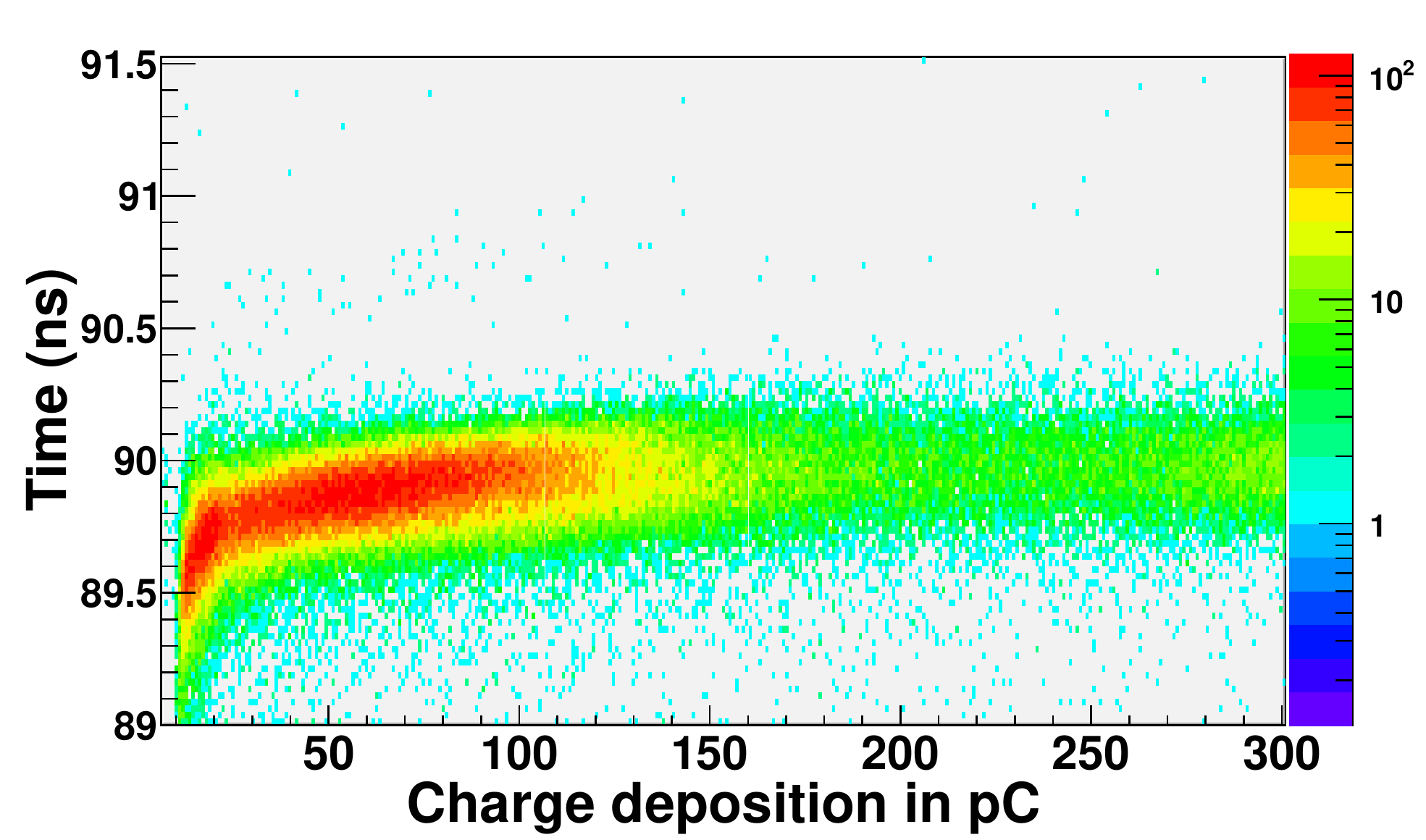}
\end{center}
\caption{Two dimensional plot (without any slew correction)for the time distribution of 
MMRPC against the charge deposited in the strip.}
\label{Raw two dimensional plot}
\end{figure}

\begin{figure}[h!]
\begin{center}
\includegraphics[width=0.7\textwidth]{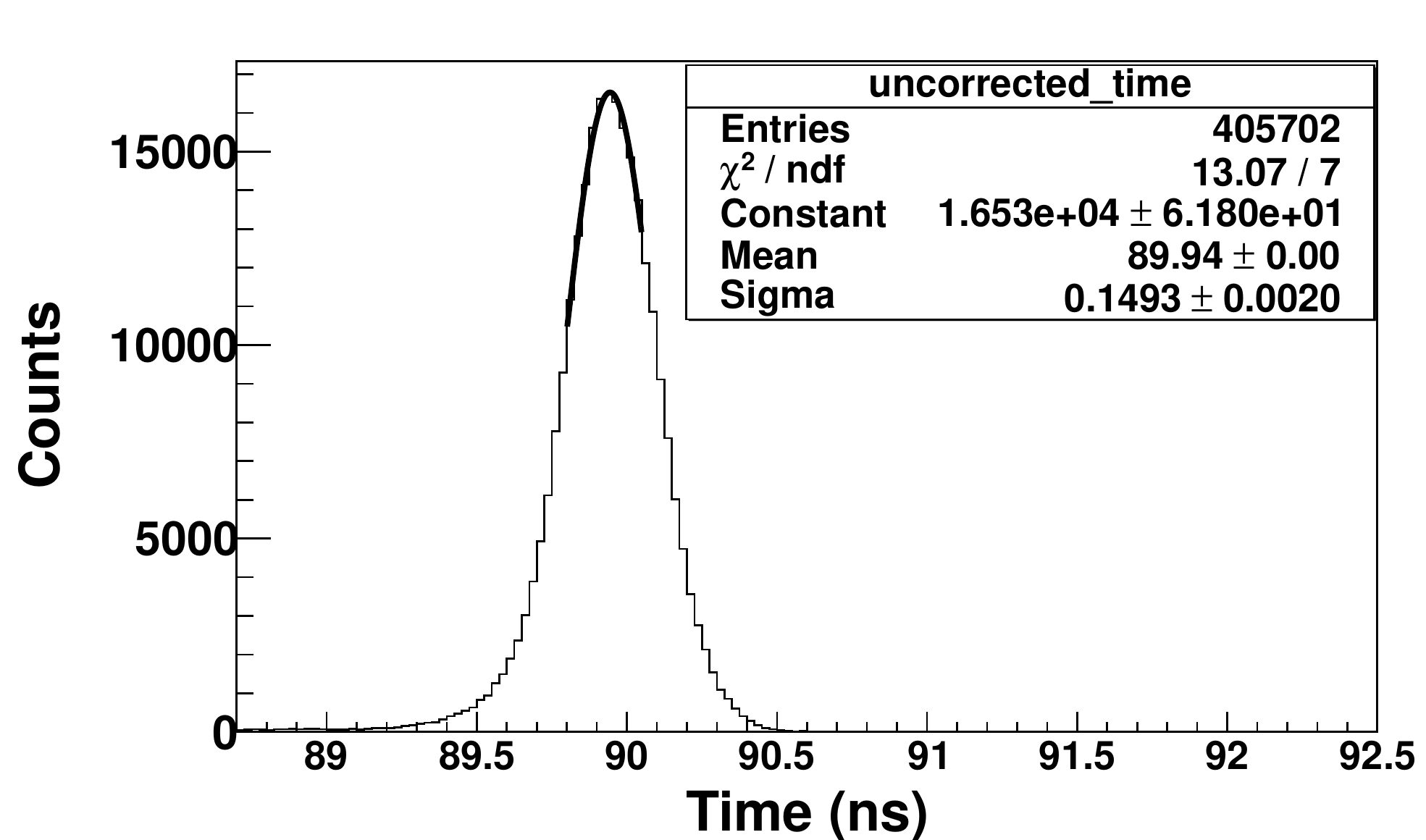}
\end{center}
\caption{Spectra for time of MMRPC without any slew correction.}
\label{Raw time resolution}
\end{figure}

\begin{figure}[ht!]
\begin{center}
\includegraphics[width=0.7\textwidth]{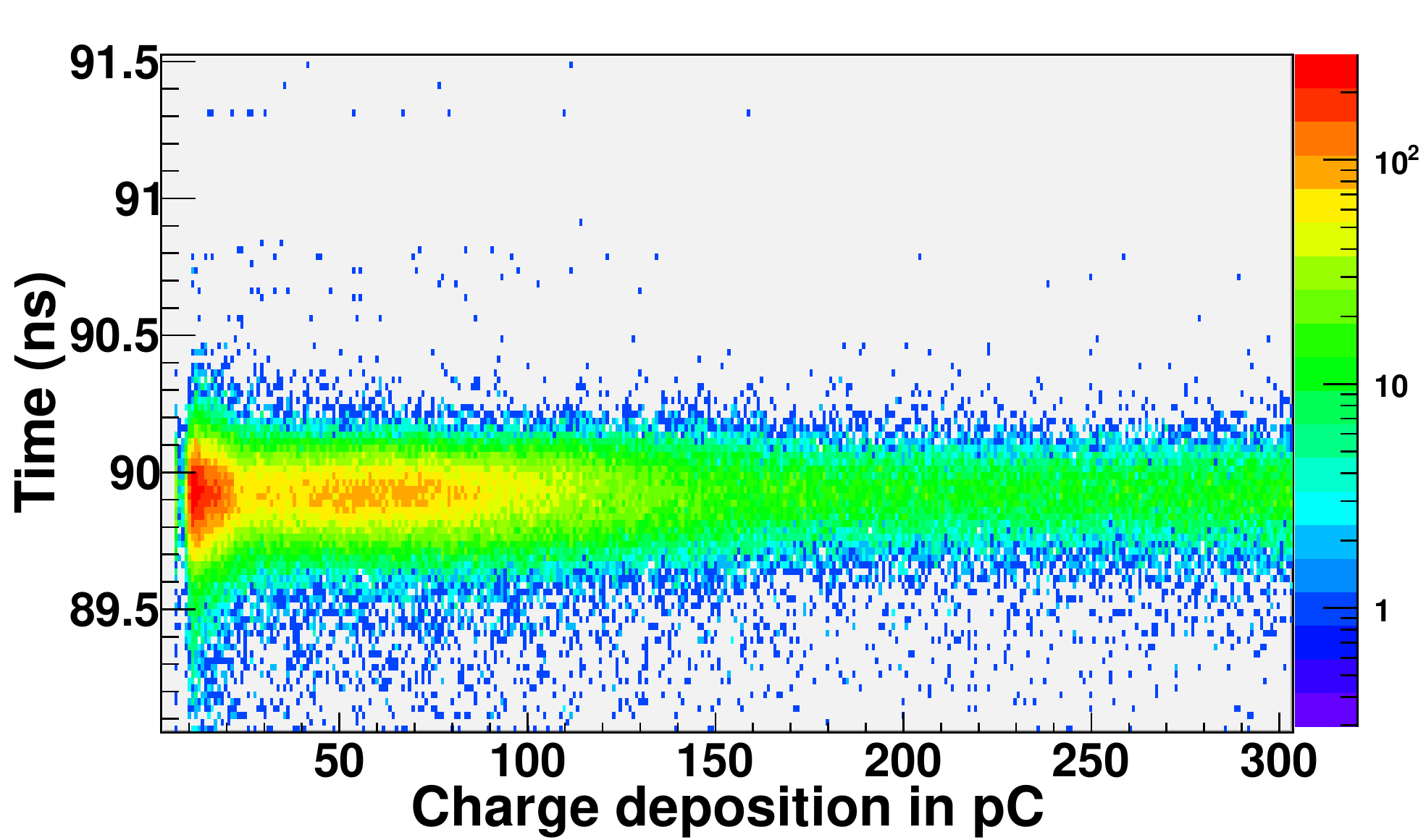}
\end{center}
\caption{Two dimensional plot for time of MMRPC after slew-correction (using our developed method) 
against deposited charge on a single strip.}
\label{Two dimensional plot after slew-correction}
\end{figure}

\begin{figure}[h!]
\begin{center}
\includegraphics[width=0.7\textwidth]{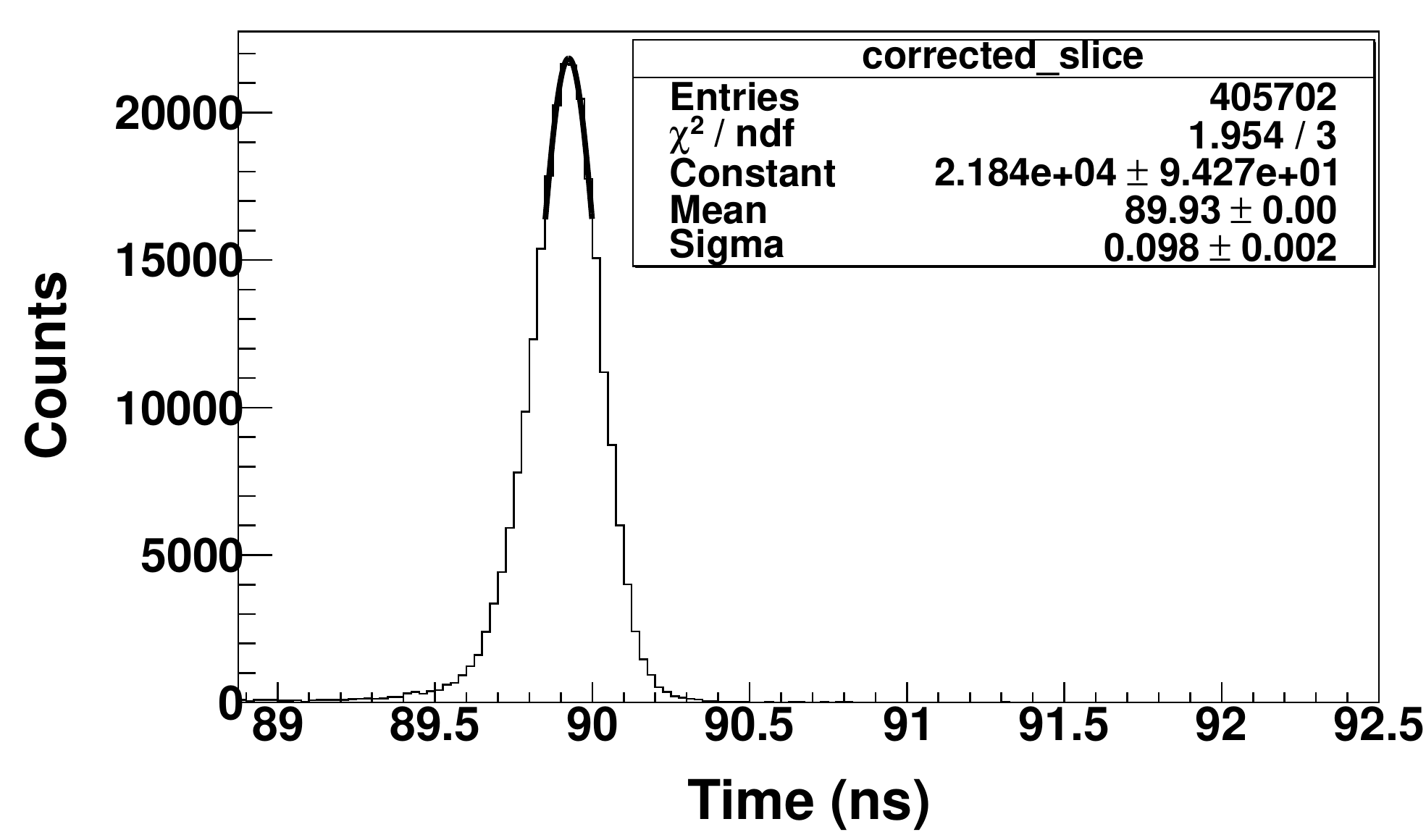}
\end{center}
\caption{Time of MMRPC after slew-correction (considering our developed method).}
\label{Time of spectra after slew-correction}
\end{figure}

\begin{figure}[ht!]
\begin{center}
\includegraphics[width=0.7\textwidth]{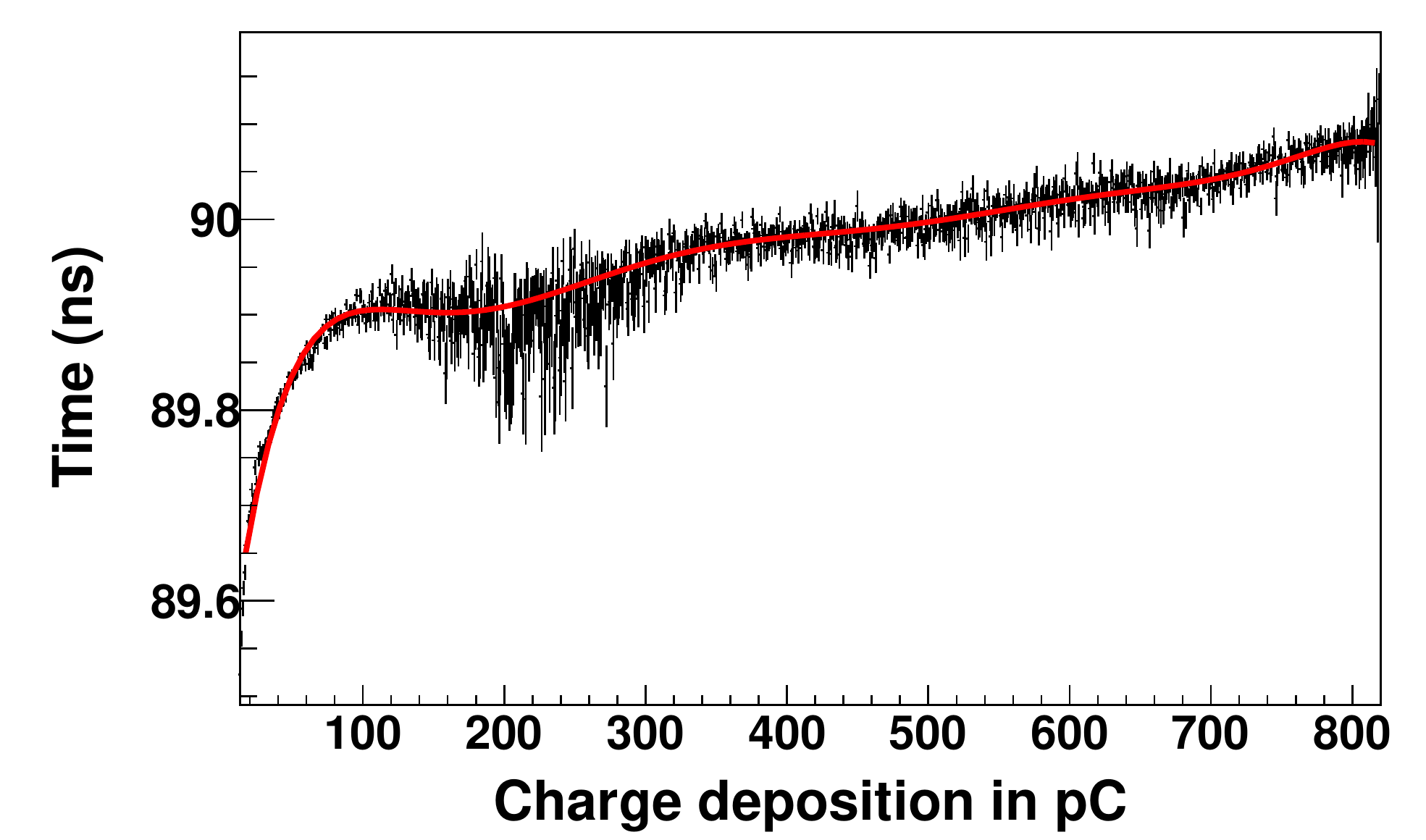}
\end{center}
\caption{2D-Profile histogram corresponding to the time distribution  of MMRPC 
(without the  slew correction) Vs.charge deposited on a single strip.}
\label{Profile histogram}
\end{figure}

Regarding the time measurements,  the mean time viz., (t$_{left}$ + t$_{right}$)/2, of the signals taken from both end of a MMRPC 
strip was considered. 
Thus time measurements, for the involved strips were obtained from the arithmetic mean of the time  obtained from the signals delivered 
from both end of a  strip of the MMRPC. Figure \ref{Raw two dimensional plot} shows a two-dimensional plot of time distribution  of the MMRPC against 
the average charge deposited on  a particular single strip. It is observed that time information for signals with low amplitude differs 
from that of the signals with higher amplitude. The   leading edge discrimination technique  has been used in extracting the time information  by the FOPI
 cards \cite{cio}. Off-line correction  of this walk in the time measurement
 is called the slew correction. The time resolution ($\sigma_t$) without the  slew correction,  as shown in Figure \ref{Raw time resolution}, is less
 than 150 ps.  This is consistent with our previous measurements using  the same detector with the cosmic muon   at SINP laboratory \cite{datt}.
The Slew correction in the time measurement was performed using two different methods - our developed method and the conventional curve fitting method 
which are in general used for timing  of MRPC.  In our  method, the QDC spectra was divided into slices of sufficient small charge distribution width.
The  measured peak positions in the time spectra were  aligned  corresponding to  each of this QDC slices through an iteration process.  
Figure \ref{Two dimensional plot after slew-correction} shows the two dimensional plot of the time distribution  of MMRPC after the  slew correction 
against the charge deposited in a single strip. The obtained time resolution ($\sigma_t$)  for all good events (as shown in Figure \ref{Time of spectra after 
slew-correction}) after this slew correction is 98.0$ \pm $3 ps. 
Considering the correction for electronics  and  reference trigger ($\sigma_t$ $\sim$35 ps) the obtained   timing resolution was 91.5 $ \pm $ 3 ps.
 In the standard  method, a profile histogram corresponding to the two dimensional plot of Figure \ref{Raw two dimensional plot} was generated.
 The profile was then fitted with a correct curve. In principle the region of the profile with lower charge deposition should be fitted with 
an exponential function while for the rest part with higher charge deposition should be fitted with a straight line. The best fitted curve was  
obtained by fitting the two region with two multi-order polynomial. Figure \ref{Profile histogram} shows the fitted two dimensional  profile histogram. 
The idea of the slew correction is to linearize the profile  through different iterations. Figure \ref{Slewed two dimensional plot} shows 
the  same 2-Dimensional plot after the slew correction
 using the conventional curve fitting method. The time resolution ($\sigma_t$) obtained in this method after slew correction is 111.3 $ \pm $3.0 ps 
(Figure \ref{Slewed time of spectra}).
The variation of time resolution (without the slew correction) against bias voltage has been shown in Figure \ref{time resolution against bias voltage}. 
A trend of improving time resolution with increasing MMRPC bias voltage was observed. The time resolution improves with increasing bias voltage and
 reaches a minima at bias voltage $\sim$ 7.5 kV. Figure \ref{time resolution against bias voltage1} shows the same plot but after the slew correction. 
The slew correction in this case was done using curve fitting method. Better time resolution of 79.6 $\pm$ 1.0 ps was  obtained by considering  of events
 from a single  strip of the detector.

The Position resolution ($\sigma_x$, $\sigma_y$) along the strip was measured from the difference in time measurement from both end of a strip. 
 The Data was considered for which the beam spot was positioned 7.5 cm and 15 cm away from the center along the length of strip-4. The
 difference in time from the two TDCs at both end of a strip reflected in the signal velocity through the anode strip. 
Considering signal velocity  (13.8 cm/ns) and time resolution (~100ps)
 the   measured position resolution ($\sigma_x$) of MMRPC along the strip was  around  2.8$ \pm $0.6 cm. 
The strip width being 2 cm, the position  resolution ($\sigma_y$) across the strip  is 0.58 cm.

\begin{figure}[h!]
\begin{center}
\includegraphics[width=0.7\textwidth]{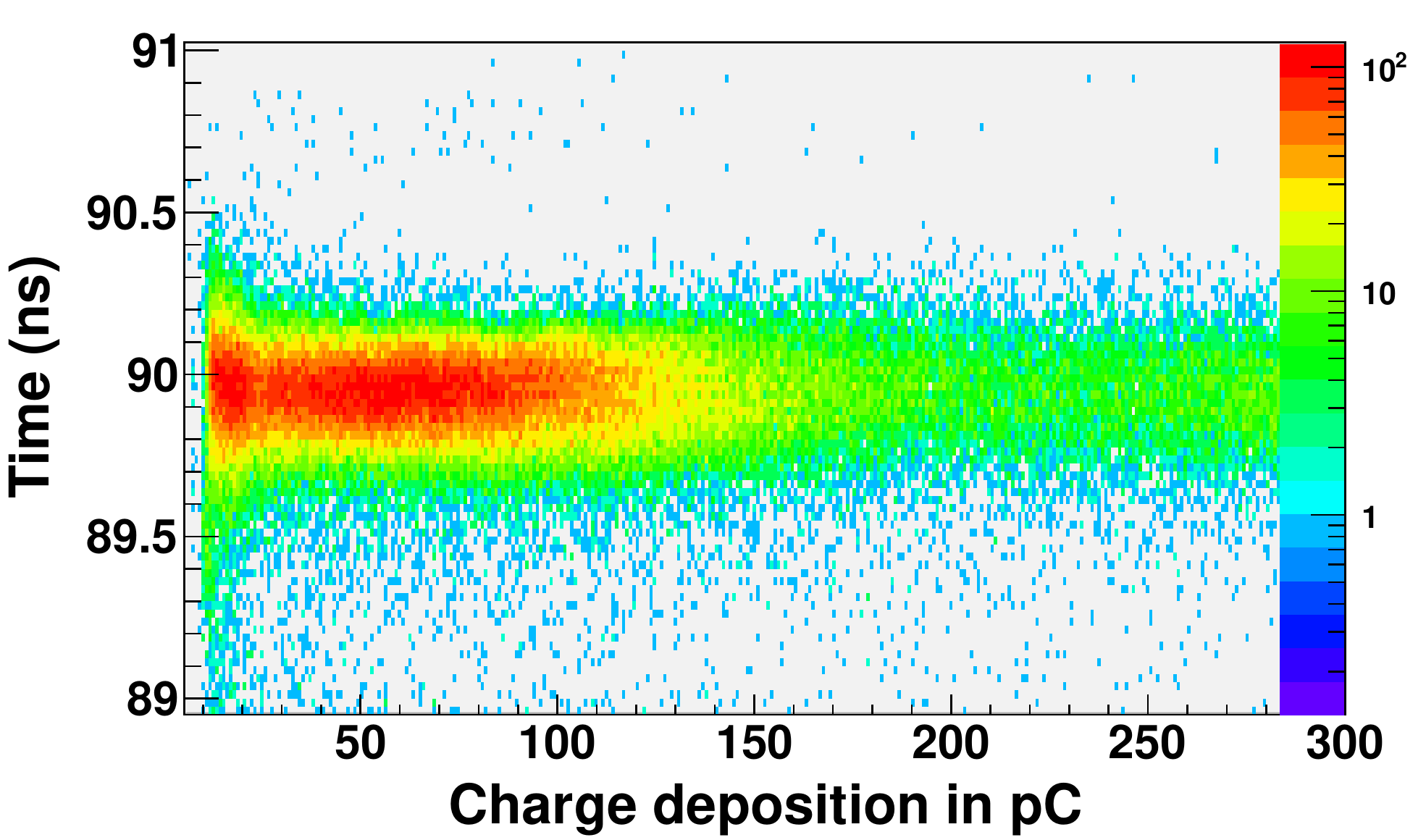}
\end{center}
\caption{Plot for time distribution of MMRPC after the slew-correction using conventional curve fitting method against 
deposited charge on a single strip.}
\label{Slewed two dimensional plot}
\end{figure}

\begin{figure}[h!]
\begin{center}
\includegraphics[width=0.7\textwidth]{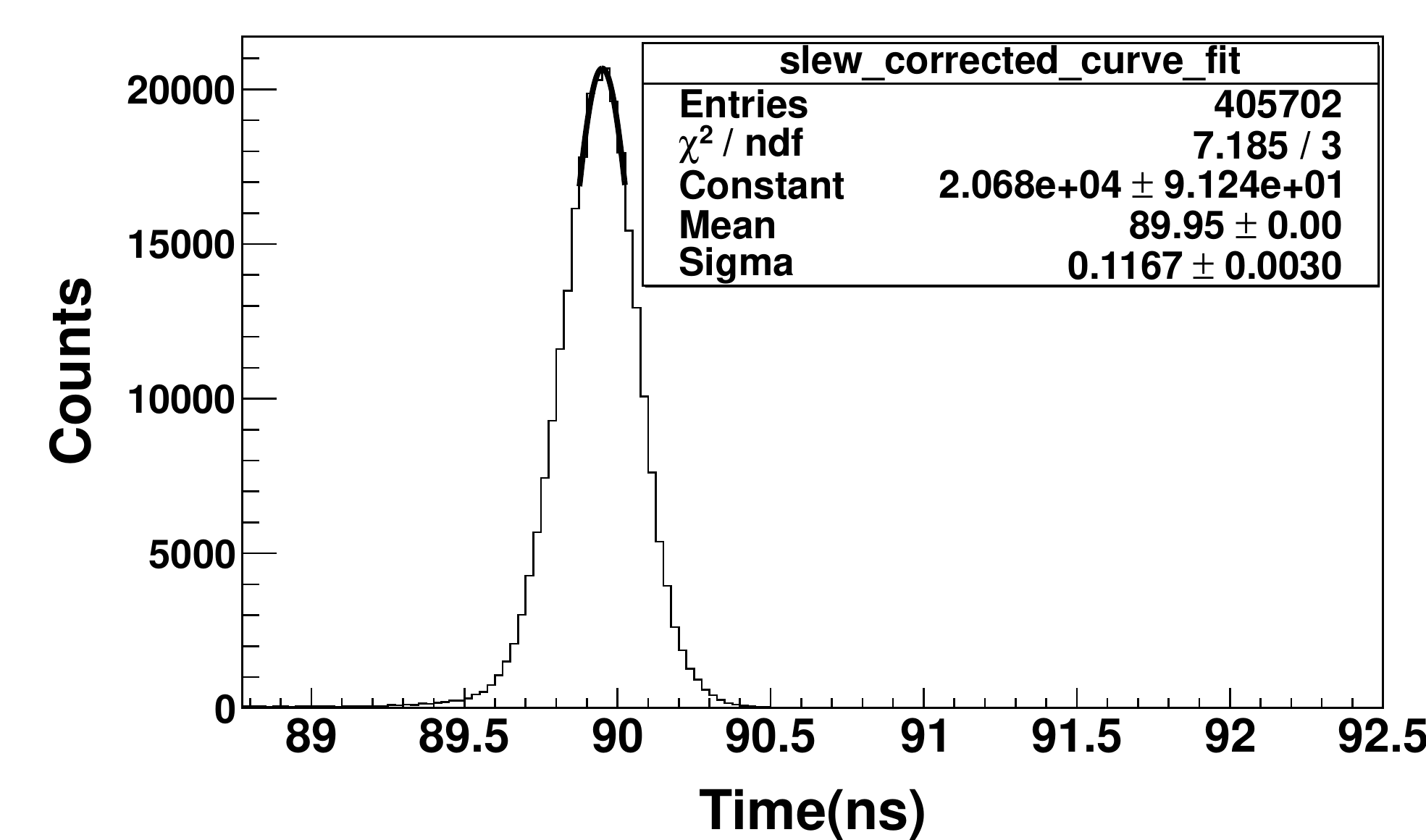}
\end{center}
\caption{Time distribution of MMRPC after the slew-correction using the curve fitting method.}
\label{Slewed time of  spectra}
\end{figure}

\begin{figure}[h!]
\begin{center}
\includegraphics[width=0.7\textwidth]{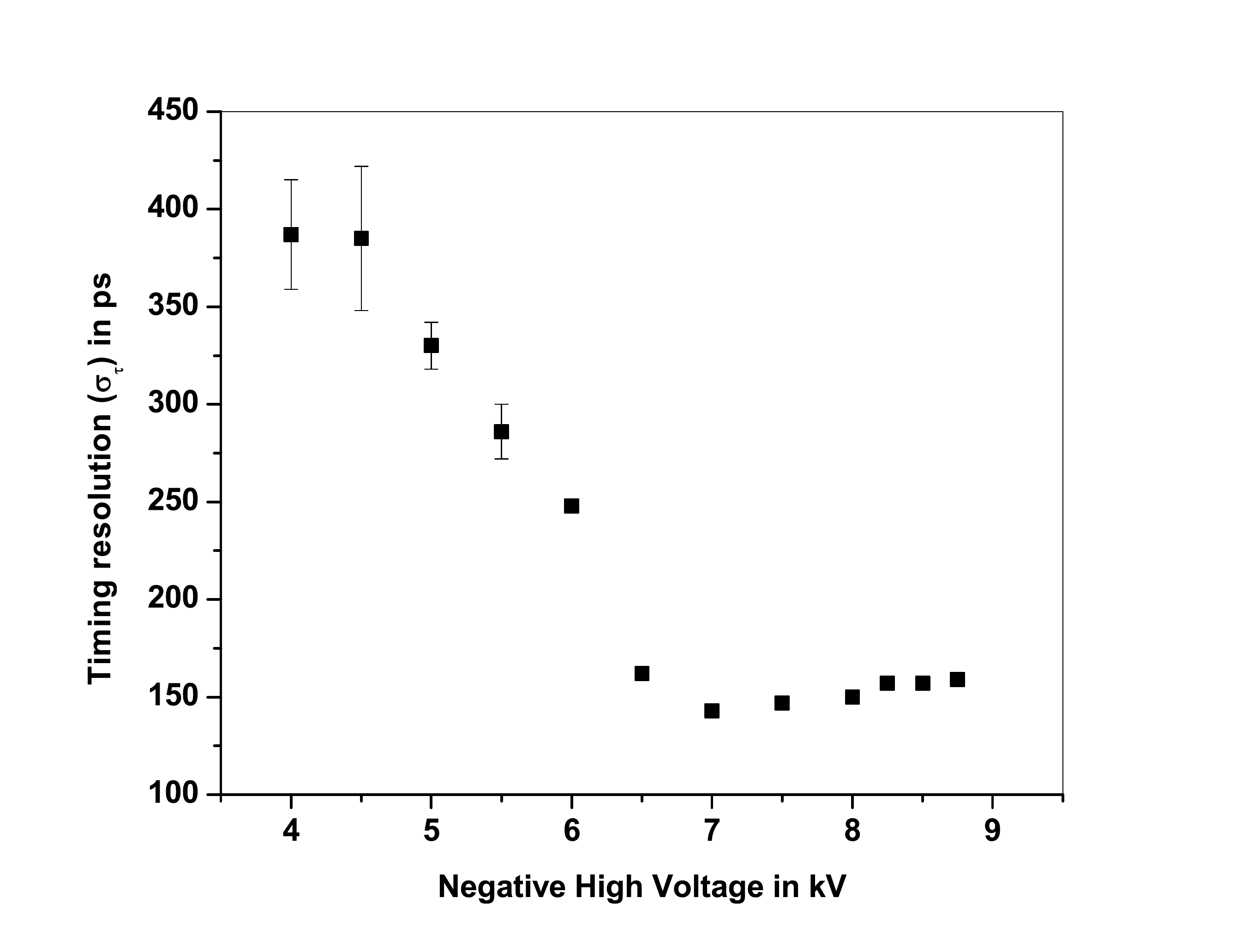}
\end{center}
\caption{Variation of the time resolution ( $\sigma_t$) of MMRPC against the bias voltage.}
\label{time resolution against bias voltage}
\end{figure}

\begin{figure}[h!]
\begin{center}
\includegraphics[width=0.7\textwidth]{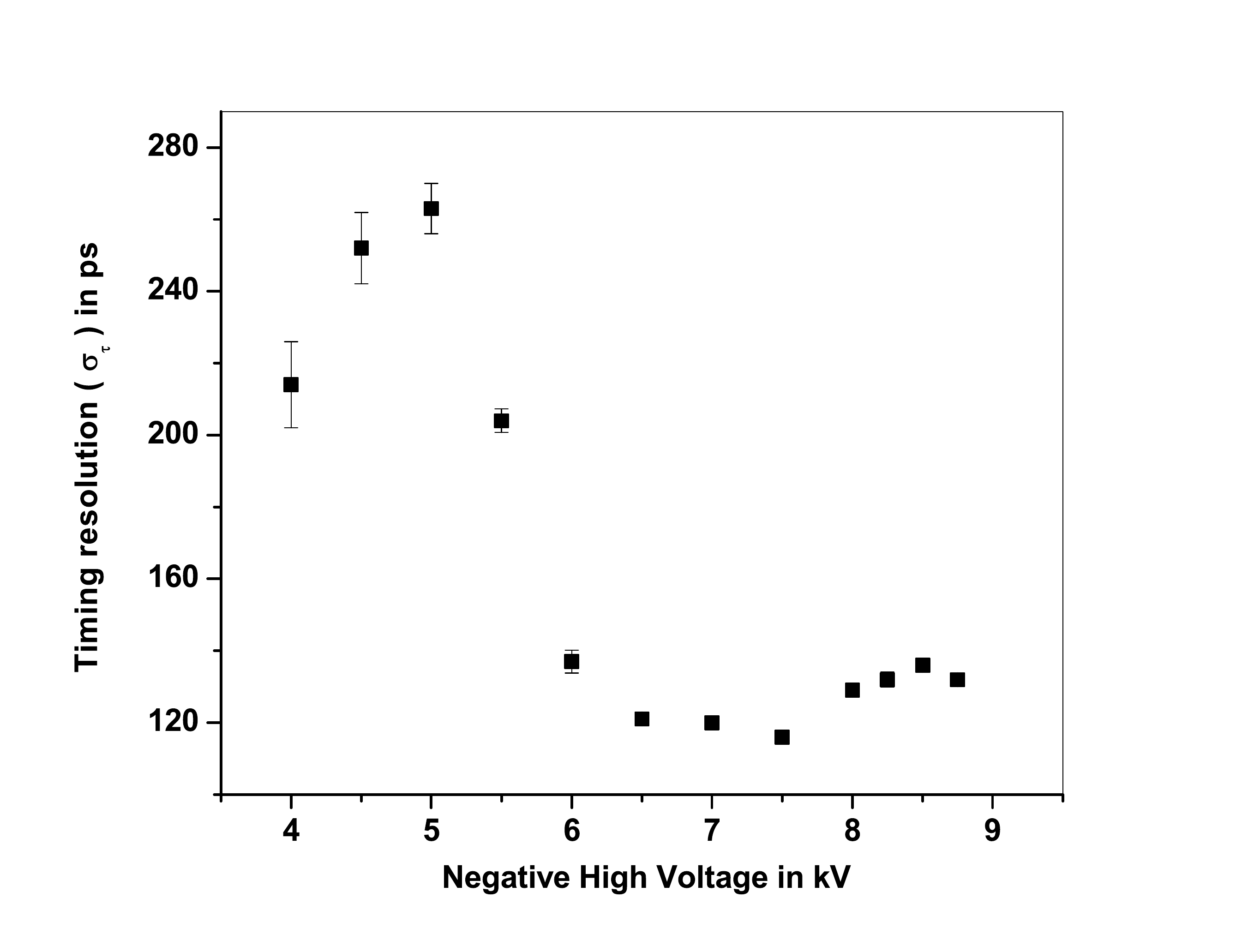}
\end{center}
\caption{Variation of the  time resolution ( $\sigma_t$)  against the bias voltage of MMRPC after the slew correction
 using the curve fitting method.}
\label{time resolution against bias voltage1}
\end{figure}

\begin{figure}[h!]
\begin{center}
\includegraphics[width=0.7\textwidth]{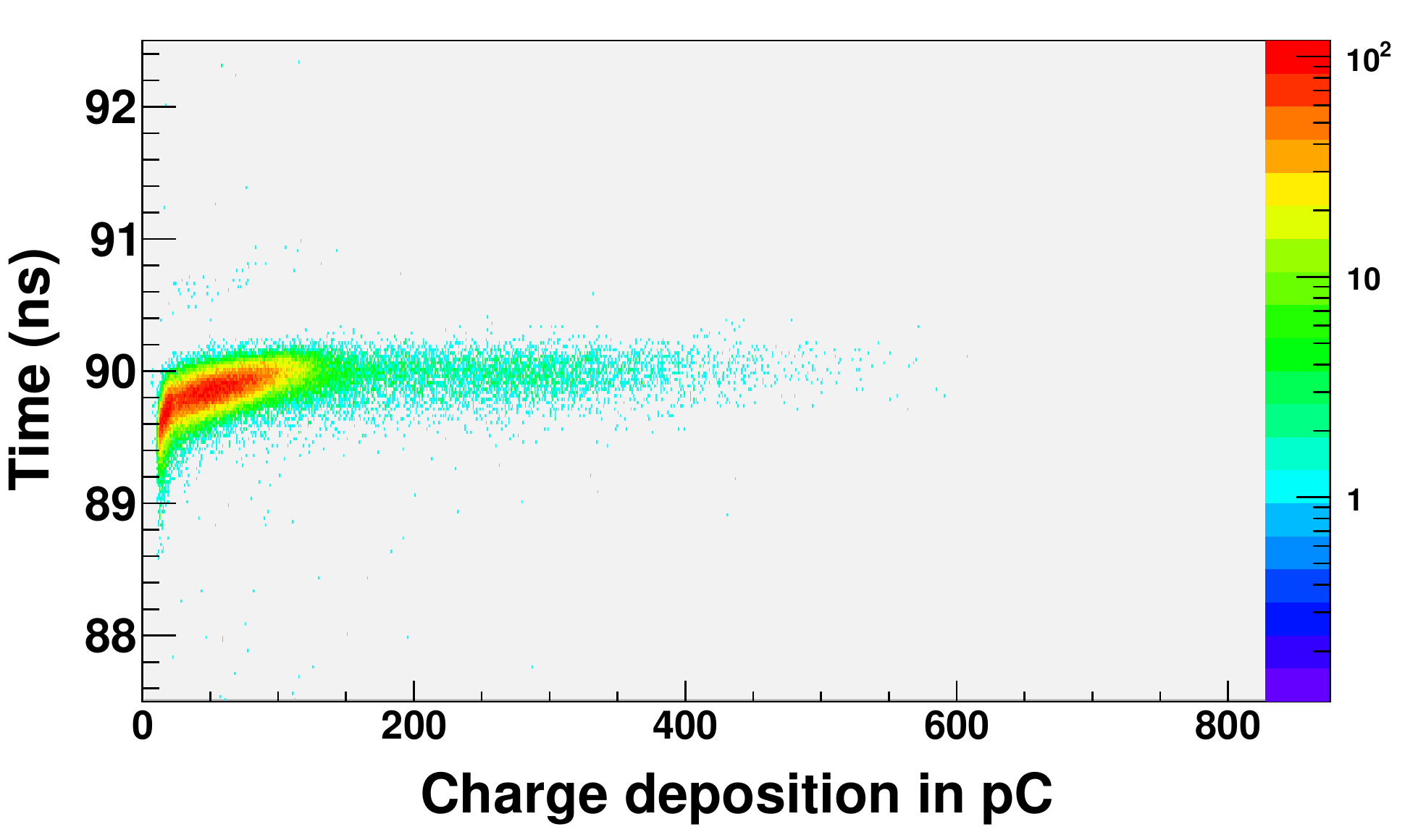}
\end{center}
\caption{Time distribution of MMRPC (without slew correction ) against the deposited charge on a strip for the events
 hitting only one strip at a time.}
\label{Two dimensional plot for only strip 4}
\end{figure}

\begin{figure}[h!]
\begin{center}
\includegraphics[width=0.7\textwidth]{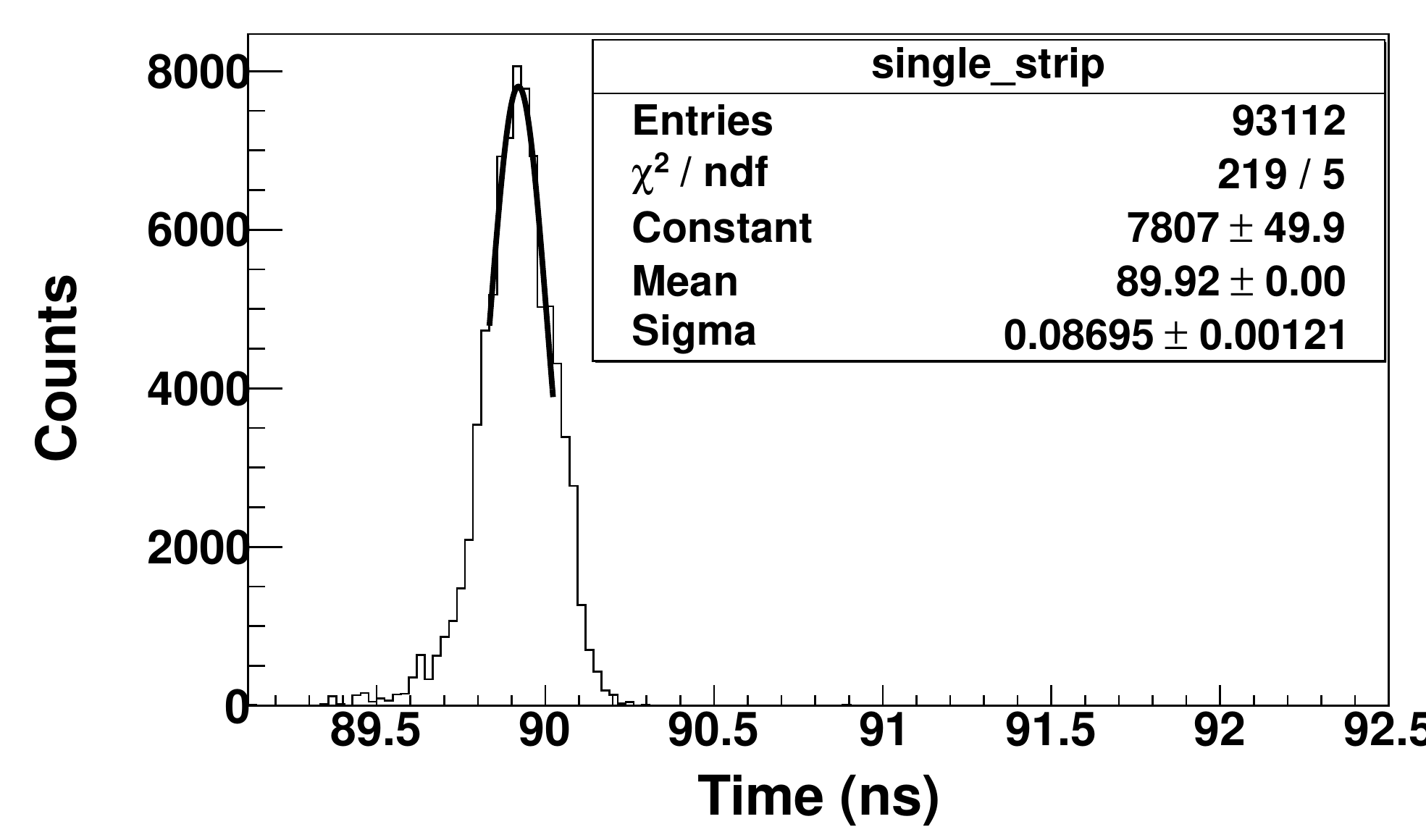}
\end{center}
\caption{Time of MMRPC for events hitting only one strip.}
\label{ToF spectra for only strip 4}
\end{figure}

\section{Discussion}

%\subsection{Subsection}
%\subsubsection{Subsubsection}

\begin{figure}
%\[h!]
\begin{center}
\includegraphics[width=0.7\textwidth]{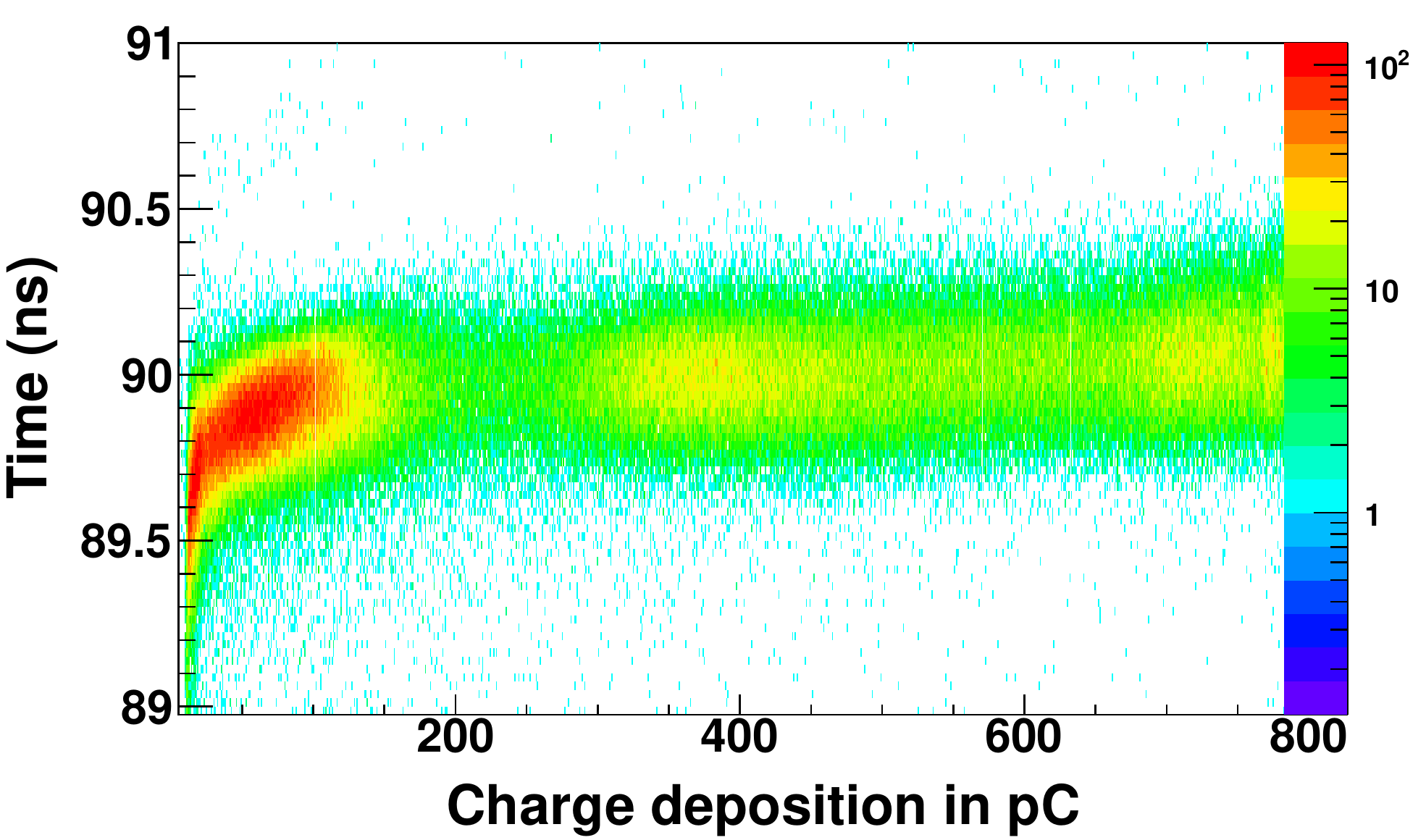}
\end{center}
\caption{Two dimensional plot for the time distribution  of MMRPC against the charge deposited  on a strip
 showing three patches at different charge deposition.}
\label{Two dimensional plot for showing three patches}
\end{figure}

\begin{figure}[h!]
\begin{center}
\includegraphics[width=0.7\textwidth]{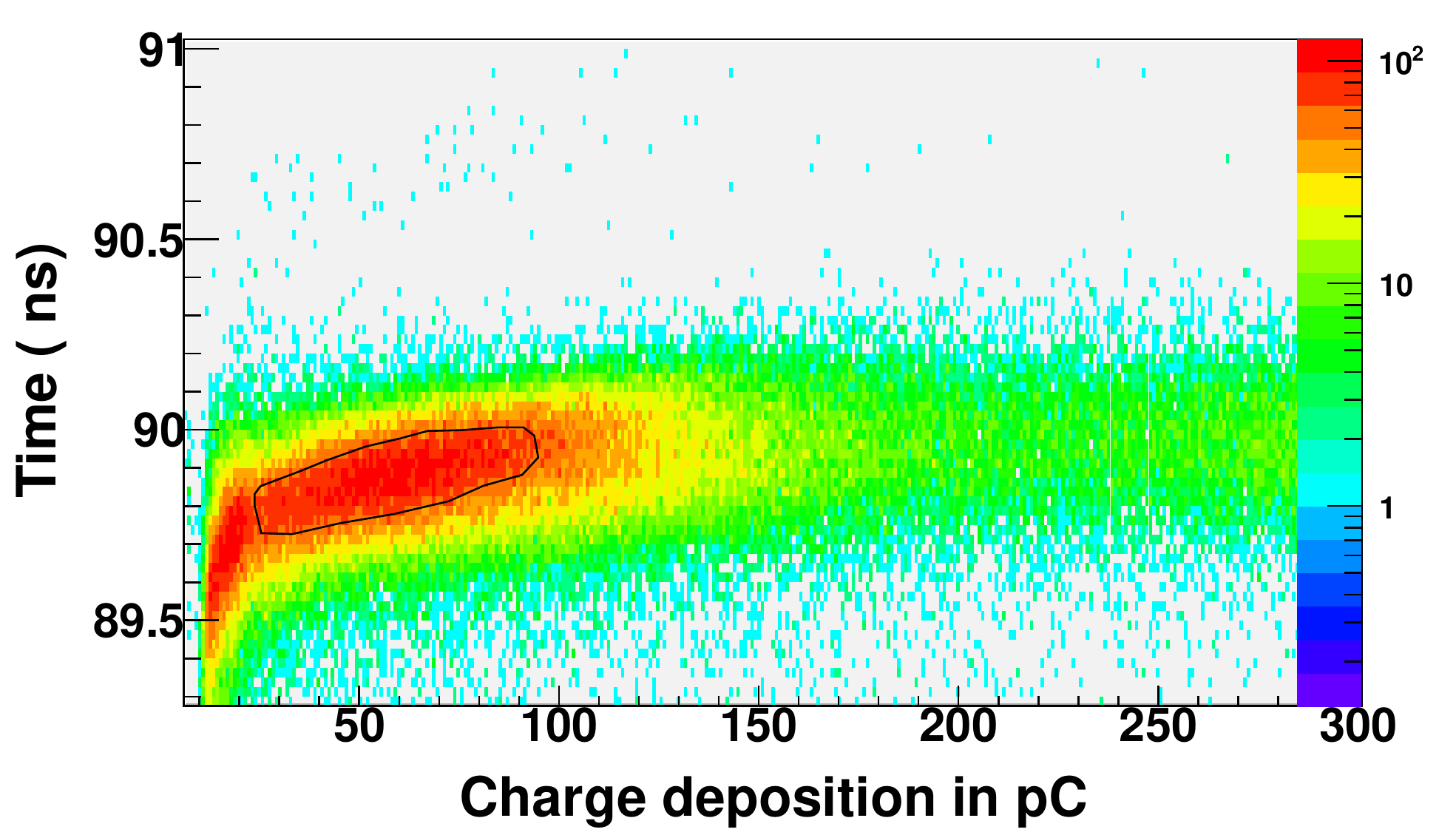}
\end{center}
\caption{Two dimensional plot for the time distribution of MMRPC against the deposited charge, showing the graphical cut.}
\label{Two dimensional plot for CUTG}
\end{figure}

\begin{figure}
%\[h!]
\begin{center}
\includegraphics[width=0.7\textwidth]{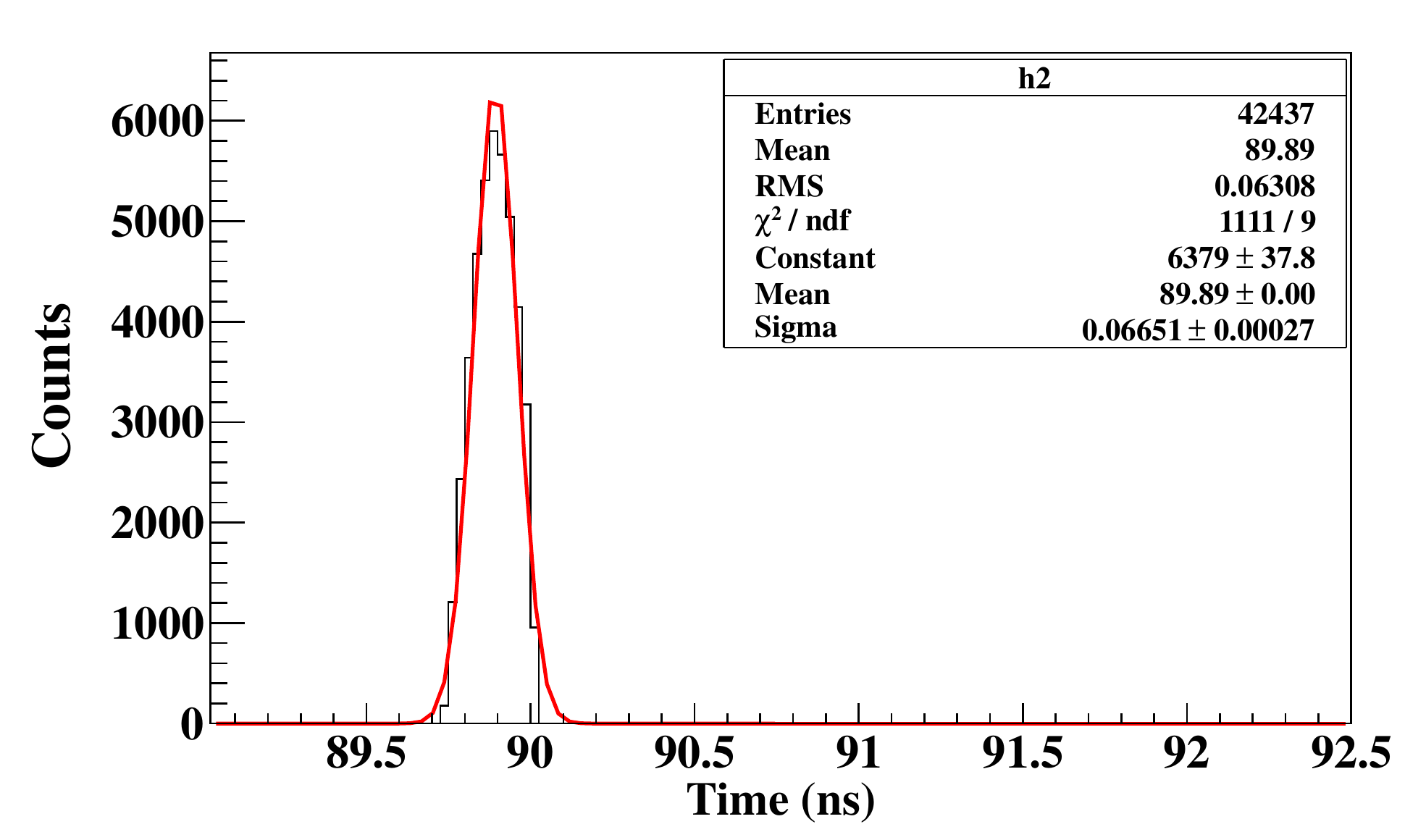}
\end{center}
\caption{Time distribution  of MMRPC corresponding to graphical cut.}
\label{ToF spectra of MMRPC for CUTG}
\end{figure}

In order to understand the operational characteristics of the MMRPC detector, the time resolution and absolute efficiency of 
the detector was measured  by considering  of events selected according to different schemes or trigger conditions. 
In one such scheme, events were selected for which only one strip has fired. 
The two dimensional plot for time distribution of MMRPC against deposited charge on the corresponding strip looked clean as shown 
in Figure \ref{Two dimensional  plot for only strip 4}. The time resolution ($\sigma_t$) corresponding to these events 
 before and after slew-correction were respectively, 129 ps and 79.6 ps.  

\begin{figure}
%\[h!]
\begin{center}
\includegraphics[width=0.7\textwidth]{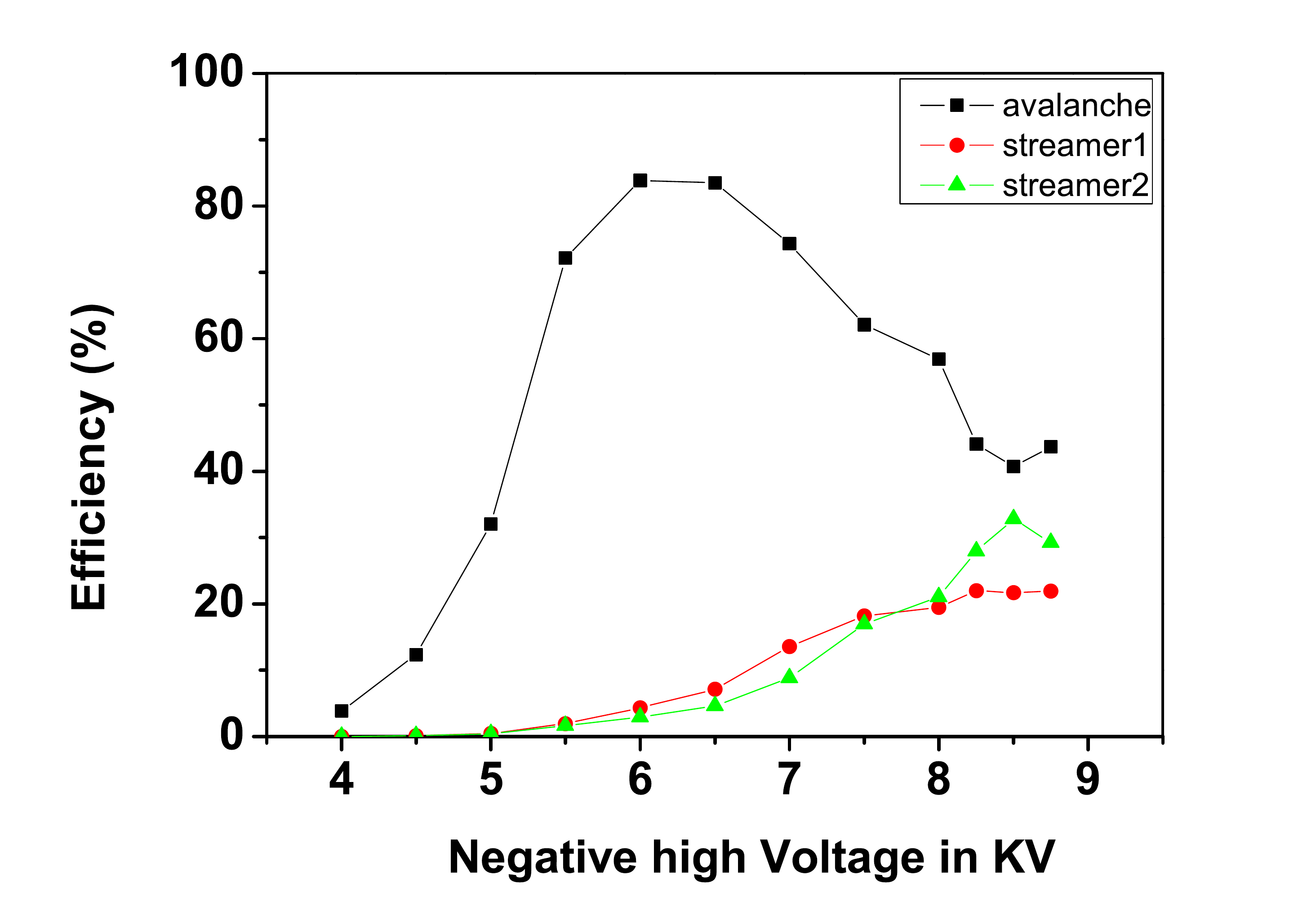}
\end{center}
\caption{Contribution of the avalanche and streamers mode of response at different bias voltage of MMRPC.}
\label{avalanche and streamers Vs bias voltage}
\end{figure}

\begin{figure}
%\[h!]
\begin{center}
\includegraphics[width=0.7\textwidth]{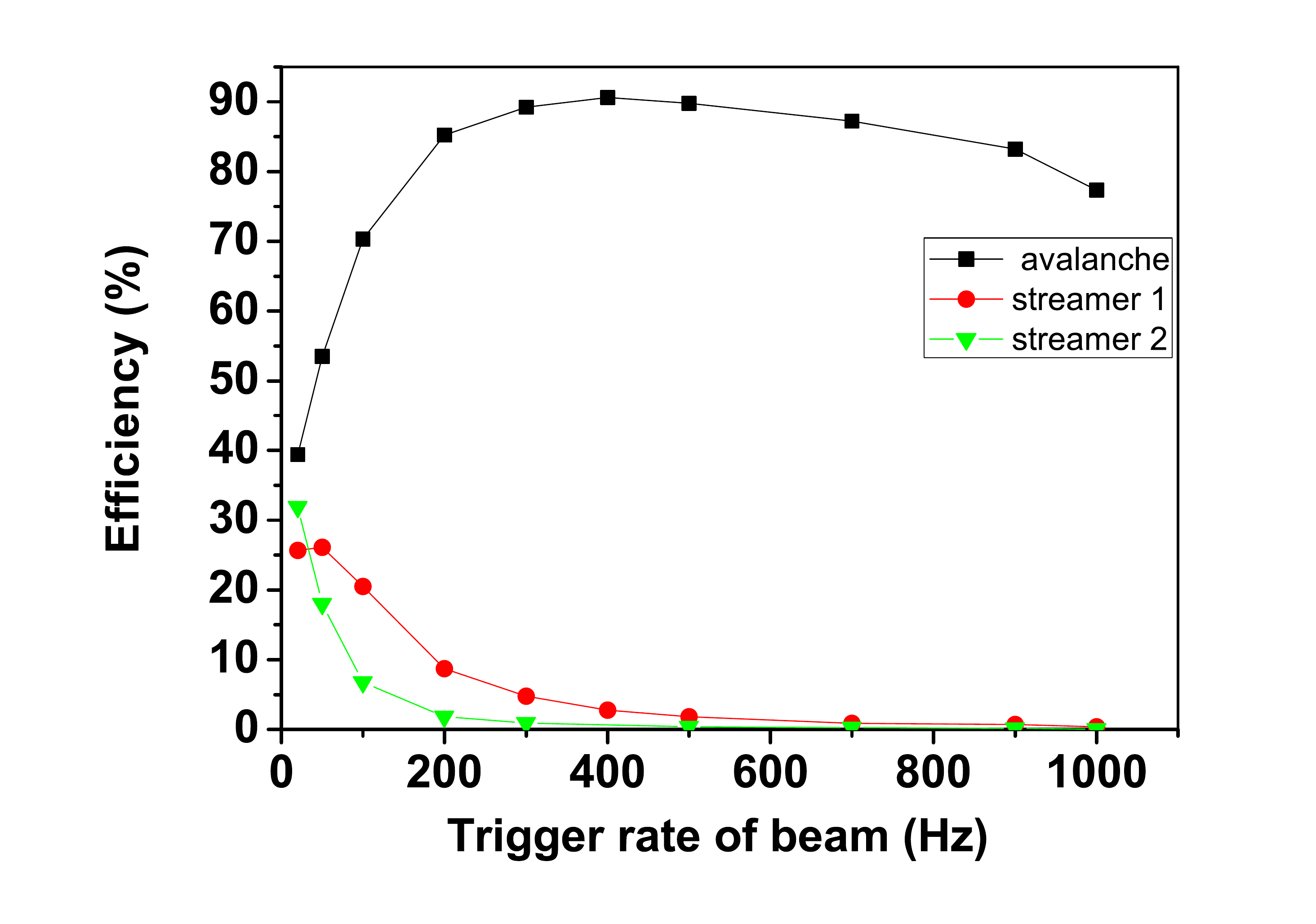}
\end{center}
\caption{Contribution of the  avalanche and streamers mode of response  at different trigger rate for a fixed bias voltage.}
\label{avalanche and streamers Vs trigger}
\end{figure}

\begin{figure}
%\[h!]
\begin{center}
\includegraphics[width=0.7\textwidth]{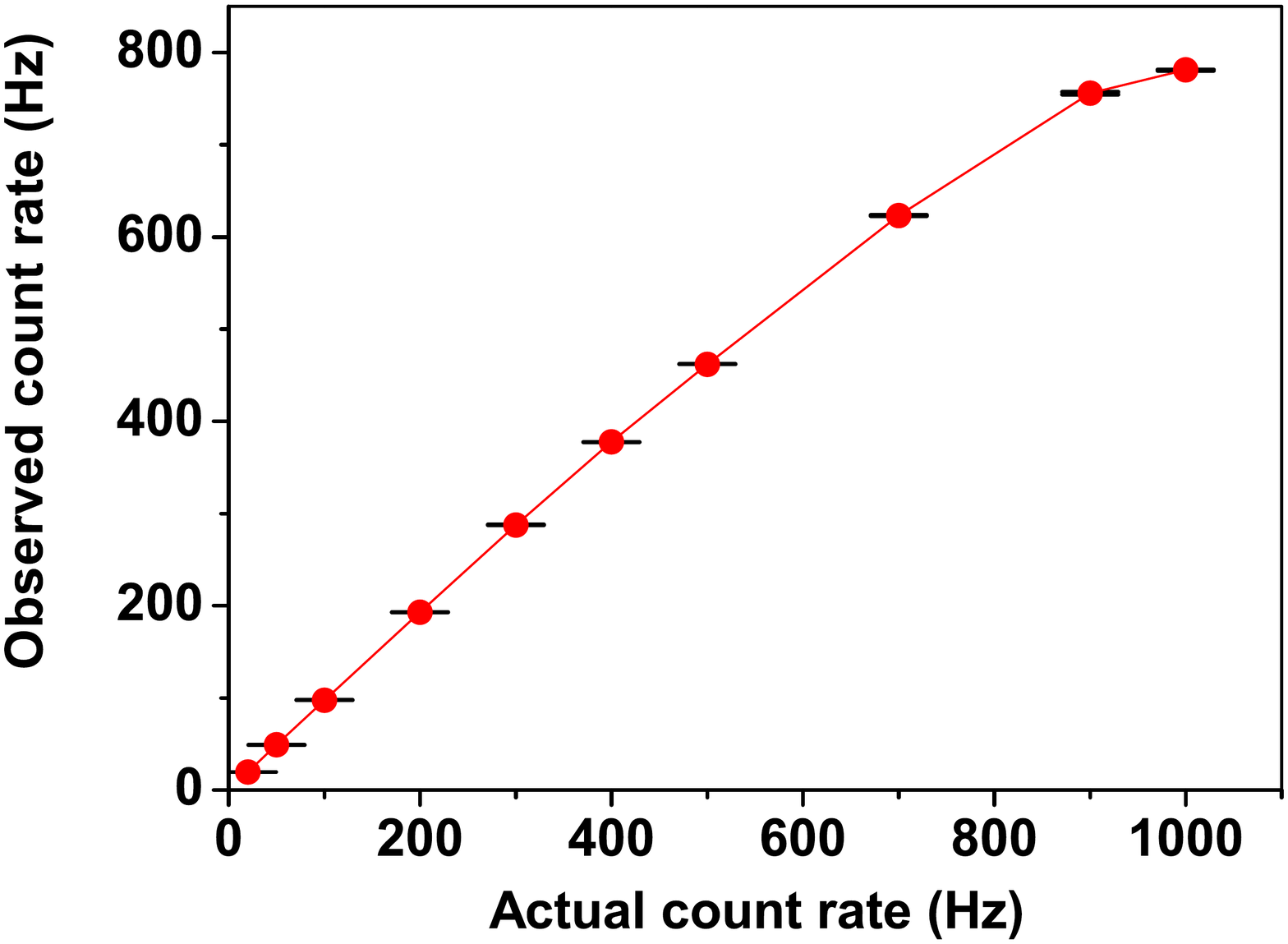}
\end{center}
\caption{Plot of observed event rate of MMRPC against actual event rate.}
\label{observed Vs actual event rate}
\end{figure}

In the two dimensional spectra for ToF of MMRPC (without slew correction) against deposited charge on a single strip,  intense
 smaller charge distribution was observed (Figure \ref{Two dimensional plot for showing three patches}). Projection 
on the time axis with a graphical cut (Figure \ref{Two dimensional plot for CUTG}) on this intense smaller charge  shows a 
time resolution ($\sigma_t$) of 56.5$\pm$1.3 ps (without slew correction) (Figure \ref{ToF spectra of MMRPC for CUTG}). 
Thus in expense of efficiency, better timing resolution can be obtained in MMRPC. This analysis shows that graphical cut
 method of analysis can be used in the analysis algorithm to avoid slew correction. Table \ref{table:Time resolution}
 shows  the time resolution and efficiency of our developed MMRPC detector with different analysis methods.

%###################################################################################################################
\begin{table*}
%\[h!]
%\begin{minipage}{5in}
\renewcommand{\thefootnote}{\thempfootnote}
\centering    

\caption{Time resolution ($\sigma_t$)   of Multi-strip Multi-gap Resistive Plate Chamber at various conditions and constaints }
%\begin{tabularx}{0.75\textwidth}{|l|X|}
\setlength{\tabcolsep}{15pt}

\begin{tabular}{l l l r r r}
\hline
  Case & Analysis methods & \multicolumn{2}{c}{Time resolution ($\sigma_t$) in ps for}  \\
\cline{3-4}
& & 1$\sigma$ fitting & 2$\sigma$ fitting & &\\
\hline

   &  &   &   & &   \\
1 & slew-correction by conventional&  111.3$ \pm $3.0 & 113.8$ \pm $1.0   \\
  & curve fitting method &   &   &  \\
  &  &   &   &  \\
2 & slew-correction by  &  91.5$ \pm $2.7 & 95.6$ \pm $0.5   \\
  & our developed method &   &   &  \\
  &  &   &   &  \\
3 & for events hitting only & 79.6$ \pm $1  & 85.2$ \pm $0.9    \\
  & strip no. 4 $\&$ slewing\footnote{slew correction done using conventional curve fitting method.} &   &    \\
  &  &   &   &  \\
4 & making graphical  & 56.5$ \pm $1.3  &    \\
  & cut before slew correction &    \\
  &  &   &   &  \\
  &  &   &   & &  \\

\hline

\end{tabular}
\label{table:Time resolution} 
%\end{minipage}
\end{table*}

%###################################################################################################################

It is interesting to note that  response mechanism of present MMRPC is mainly of  avalanche mode but in the two dimensional
 plot for time distribution  of  MMRPC against deposited charge on a single strip,   three patches at lower, intermediate and higher
 charge depositions  can be  observed (Figure \ref{Two dimensional plot for showing three patches}). In order to 
understand the causes of these charge distributions, three  different graphical cuts were made corresponding to 
the three different charge distribution.  Relative contributions of the three patches were determined from these 
three cuts. That was continued for data corresponding to different bias voltages of MMRPC. 
Figure \ref{avalanche and streamers Vs bias voltage} shows the efficiency of the three patches with  increasing 
bias voltage. It was observed that the contribution for the first patch at lower charge deposition dominates at lower
 bias voltage. It  reaches a maxima at $\sim$ 6.5KV and then decreases. The patch at lower charge deposition may be
 due to the avalanche mechanism in the detector. The contribution due to the second and third patches increases with 
increasing bias voltage. These two patches at higher charge deposition  may be due to two different streamers 
dominating at higher electric field. The efficiency of the three patches at different charge deposition was also 
 studied with the variation in trigger rate of the beam. Figure \ref{avalanche and streamers Vs trigger} shows the 
contributions of different operational mechanism  of the MMRPC detector at different event rate for a fixed bias 
voltage. It is interesting to note that at  higher event rates  streamer mode of operation  decreases where as avalanche 
mode of operation  dominates for a particular bias voltage. Figure \ref{observed Vs actual event rate} shows a plot of 
the observed  event rate in MMRPC against the actual event rate. As seen from figure 25 the system does not show any 
significant dead time upto 800 Hz. Recently, it has been shown by several groups \cite{hadd,jing} that
high  detection efficiency  with good timing resolution ( around 90 \%) can be obtained  for  response  of electrons  
with  higher events rate ( more than few  kHz)  using  MRPC  with low resistivity doped  glass.
      
\section{Summary}
A prototype of Multi-strip  Multi-gap Resistive Plate chamber (MMRPC) with active area 40 cm $\times$ 20 cm has been developed 
at SINP, Kolkata.  Detailed response of the developed detector was studied using the pulsed electron beam from ELBE 
at Helmholtz-Zentrum Dresden-Rossendorf. The response of SINP developed MMRPC with  different controlling
 parameters has been described in details. The obtained time resolution ($\sigma_t$) of the detector after slew correction 
was 91.5$ \pm $3 ps. Position resolution measured along ($\sigma_x$) and across ($\sigma_y$) the strip was 2.8$ \pm $0.6 cm 
and 0.58 cm, respectively. The measured absolute efficiency of the detector for minimum ionizing particle like electron was 
95.8$\pm$1.3 $\%$. Better  timing resolution of the detector  of 79.6$ \pm $3.5 ps  was obtained by restricting to the events 
hitting only a single strip. It was further observed that under special consideration of events using graphical cut at smaller 
charge distribution, slew correction in time resolution becomes non-essential. The efficiency corresponding to these events 
though decreases   but timing resolution as good as 53.2$ \pm $1.3 ps can be obtained. The response of the detector was mainly
 in avalanche mode but a few percentage of streamer mode  response  was also observed. 
        
\acknowledgments
The development of the detector fully  funded by the XIth plan project, SEND project (PIN:11-R\&D-SIN-5.11-0400),
Dept. of Atomic Energy (DAE), Govt. of India. Authors are deeply thankful to the members of workshop of Saha Institute of
 Nuclear Physics for building various components of detector and to the accelerator  people of ELBE, Dresden  for their
 support during the experiment. Authors are thankful to Prof. Sudeb Bhattacharya, Kolkata for critically reviewing  the manuscript
and providing suggestions.

\end{document}